\documentclass[sigplan,nonacm]{acmart}

\usepackage[utf8]{inputenc}
\usepackage{graphicx}
\usepackage{hyperref}
\usepackage{xspace}
\usepackage{listings}
\usepackage{xcolor}
\usepackage{array}

\copyrightyear{2025}
\acmYear{2025}
\setcopyright{acmlicensed}
\acmConference[SYSTOR '25]{The 18th ACM International Systems and Storage Conference}{September 8--9, 2025}{Virtual, Israel}
\acmBooktitle{The 18th ACM International Systems and Storage Conference (SYSTOR '25), September 8--9, 2025, Virtual, Israel}
\acmDOI{10.1145/3757347.3759129}
\acmISBN{979-8-4007-2119-9/2025/09}

\begin{document}

\newcommand{\little}{{\em little3}\xspace}
\newcommand{\hyang}{{\em hyang5}\xspace}
\newcommand{\gates}{{\em gates3}\xspace}
\newcommand{\littlehyang}{{\em little3 + hyang5}\xspace}
\newcommand{\littlehyanggates}{{\em little3 + hyang5 + gates3}\xspace}

\newcommand{\itemParagraph}[1]{\vspace{0.3em}\noindent\textbf{#1}}
\newcommand{\afkey}{{affinity key}\xspace}
\newcommand{\afkeys}{{affinity keys}\xspace}
\newcommand{\partkey}{{partition key}\xspace}
\newcommand{\partkeys}{{partition keys}\xspace}
\newcommand{\hashtag}{{hash tag}\xspace}
\newcommand{\hashtags}{{hash tags}\xspace}
\newcommand{\afgrouping}{{affinity grouping}\xspace}
\newcommand{\Afgrouping}{{Affinity grouping}\xspace}
\newcommand{\afgroup}{{affinity group}\xspace}
\newcommand{\afgroups}{{affinity groups}\xspace}

\newcommand{\objectpool}{{ object pool}\xspace}
\newcommand{\objectpools}{{object pools}\xspace}

\title{Keep Your Friends Close: Leveraging Affinity Groups\\to Accelerate AI Inference Workflows}
\renewcommand{\shorttitle}{Keep Your Friends Close: Leveraging Affinity Groups to Accelerate AI Inference Workflows}

\author{Thiago Garrett}
\orcid{0000-0002-7171-5463}
\affiliation{%
    \institution{University of Oslo}
    \city{Oslo}
    \country{Norway}
}
\email{thiagoga@ifi.uio.no}

\author{Weijia Song}
\affiliation{%
    \institution{Cornell University}
    \city{Ithaca}
    \country{USA}
}
\email{ws393@cornell.edu}

\author{Roman Vitenberg}
\affiliation{%
    \institution{University of Oslo}
    \city{Oslo}
    \country{Norway}
}
\email{romanvi@ifi.uio.no}

\author{Ken Birman}
\affiliation{%
    \institution{Cornell University}
    \city{Ithaca}
    \country{USA}
}
\email{ken@cs.cornell.edu}

\renewcommand{\shortauthors}{Garrett et al.}

\begin{abstract}%
AI inference  workflows are typically structured as a pipeline or graph of AI programs triggered by
events.  As events occur, the AIs perform inference or classification tasks under time pressure
to respond or take some action.  Standard techniques that reduce latency in
other streaming settings (such as caching and optimization-driven scheduling)
are of limited value because AI data access patterns (models, databases) 
change depending on the triggering event: a significant departure from traditional streaming.  In this work, we propose
a novel \afgrouping mechanism that makes it easier for developers to express
application-specific data access correlations, enabling coordinated management
of data objects in server clusters hosting streaming inference tasks. Our proposals are thus complementary to other approaches such as caching and scheduling.
Experiments confirm the
limitations of standard techniques, while showing that the proposed mechanism is able to maintain
significantly lower latency as workload and scale-out increase, and yet requires
only minor code changes.

\end{abstract}




\maketitle

\section{Introduction}\label{sec:intro}

Latency-sensitive AI-based applications are increasingly common~\cite{olda}. For instance, in edge
intelligence applications, devices pass captured data through AI inference and classification
pipelines, frequently combined with continuous learning~\cite{edgeint}.  Minimizing latency is the
priority, although throughput and resource utilization remain important goals~\cite{clipper}.  

Our work starts with the observation that when these pipelines are deployed onto standard Stream
Processing (SP) platforms~\cite{stream}, a mismatch occurs.  SP systems are typically optimized for
highly parallel stateless computations and data transformations. AI inference and classifiers are
often stateful and context-sensitive, resulting in data dependencies that can defeat the assumptions
made in SP schedulers~\cite{edgeserve}.

The issue is further complicated by unpredictability.  For example, a trajectory-computing task for
mobile ``actors'' in a traffic scene might have different models trained for cars, bicycles,
pedestrians, etc.  Among the vehicle models there could be one specialized for taxis, another for
emergency responders, and a general-purpose model for all others.  The AI won't know which actors
are present in the scene (and hence which objects will be needed) until runtime.  Caches are too
small to hold a copy of every object that might be needed, yet even a single cache miss will leave
the task waiting while data is fetched over the network.

Data movement overheads are further amplified when different stages of an AI pipeline are executed
in different nodes of a network. Reasons for distributing different stages across multiple nodes
include scalability, load balancing, parallelism, and specialized hardware sharing. Scalable storage
tends to randomly distribute data objects across the infrastructure. Although schedulers try to
minimize data movement when deciding where to execute a pipeline stage, there is no collocation
guarantee.

AI developers understand well which collocation dependencies would be beneficial for the latency of their applications. Alas, existing platforms (such as standard cloud frameworks) lack the hooks required for developers to provide feedback. We conjecture that this is partly because of a presumption that such a mechanism would be in tension
with core platform features such as load balancing and auto-scaling, and in part because of the concern that the annotation would require significant effort from the developers.


In this work, we propose the {\em \afgrouping} mechanism, which has two main goals: (i) to offer
standardized, platform-independent annotation method whereby developers can express application-specific
knowledge about data/computation correlations in a deployment-agnostic way (allowing 
mechanisms such as auto-scaling and load balancing to keep doing their job); and (ii) leverage this
data to significantly reduce latency as workload increases and the platform scales out.  The
mechanism consists of three elements: developer-specified logic that attaches a string, called an
{\em \afkey}, to each new incoming request; developer-specific code to tag stored data objects with
\afkeys; and a runtime engine that makes use of \afkeys to optimize object and task placement
decisions.  Leveraging affinity groups does require changes to the runtime platform, but involves
minimal changes to the AI application. Benefits include enabling proactive collocation of data with
computation, coordinated prefetching, and more efficient cache management. 
Our proposals are thus complementary to other approaches such as caching and scheduling.

In order to evaluate our proposals, we design and implement a representative latency-sensitive
AI-based application consisting of a composition of off-the-shelf AI models. Using this application,
two sets of experiments are performed. First, we deploy our representative application on a
state-of-the-art stream processing platform targeted at latency-sensitive AI workflows, which is
based on a K/V store. Results show that \afgrouping achieves significantly lower and more consistent
latency as workload and scale-out increase, compared to the standard object and task placement
strategy of the employed platform. All this requires a minimal annotation effort from the developer.

In the second set of experiments, we deploy the same application on a public cloud, Microsoft Azure.
The goal is to show evidence of the data movement overheads present in existing SP and AI serving
platforms. We observe frequent pipeline stalls stemming from data accesses that required fetching
objects over the network. We then modify our application to address the observed overheads, at the
cost of a high coupling between application and deployment.

\begin{figure}
  \centering
  \includegraphics[width=0.6\linewidth]{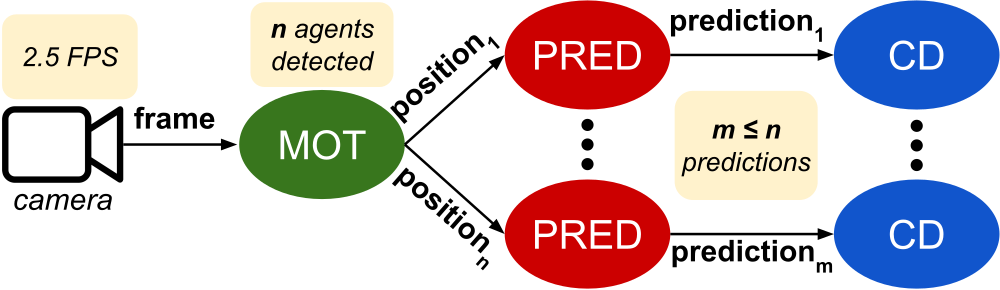}
  \caption{Computational graph of the RCP application.}
  \Description{Computational graph of the RCP application.}
  \label{fig:app_graph}
  \vspace{-5mm}
\end{figure}

This work makes the following contributions:
\begin{itemize}
\item We identify inefficiencies of modern platforms to collocate data when deploying a representative latency-sensitive AI-based application that comprises off-the-shelf AI models.
We support this claim by investigating the application performance on a public cloud.
\item To address the identified issue, we propose \afgrouping, an easy-to-use mechanism that requires no developer knowledge of deployment/environment details, and yet enables platforms to achieve effective collocation of data and compute.
\item An implementation of the \afgrouping mechanism on a state-of-the-art AI serving platform based on a K/V store is described and evaluated.
\end{itemize}

The remainder of this work is organized as follows.  Section~\ref{sec:motivation} motivate our
proposals by introducing a representative latency-sensitive AI-based application and identifying
shortcomings of modern platforms in face of the application data access patterns. This application
is used as a running example throughout this paper.  We then propose the \afgrouping mechanism in
Section~\ref{sec:affinity}.  Experimental results showing the benefits of the proposed \afgrouping
mechanism on a local cluster are reported in Section~\ref{sec:cascade}.  A series of experiments on
Microsoft Azure that support our motivating claims 
are reported in Section~\ref{sec:azure}.  Related work is presented in Section~\ref{sec:related}, while Section~\ref{sec:beyond} discuss the generality of our approach.
We conclude the paper in Section~\ref{sec:conc}.

\section{Motivation}\label{sec:motivation}

We first introduce a Real-Time Collision Prediction (RCP)
application that is used as a running example in this paper.
RCP is representative of a large, important class of latency-sensitive
applications~\cite{inferline} seen in settings such as factories, robotic
warehouses and commercial retail. We then identify limitations of modern SP and
AI serving platforms in face of the data access patterns showcased by the RCP
application. 

\subsection{Real-Time Collision Prediction}
\label{subsec:application}

The RCP application continuously sense imminent
collisions between actors (e.g. cars, pedestrians) in traffic intersections with
a window of a few seconds. It consists of a composition of off-the-shelf AI
models (trained for different tasks), resulting in a pipeline that runs in an
edge datacenter, near traffic cameras. We focus on the AI inference
pipeline, but examples of reactions to the pipeline output could include flashing red
lights or notifying suitably equipped vehicles.  As seen in
Figure~\ref{fig:app_graph} the RCP pipeline has three steps: multi-object
tracking (MOT), trajectory prediction (PRED), and collision detection (CD).
Implementation details are provided in Section~\ref{subsec:app_impl}.     


The first step, MOT, is responsible for detecting and tracking actors in video
frames received from cameras.  The computation in this step consists of
detecting all actors in each frame and their positions, and then matching each
detected actor with its prior instance(s) if any, to determine speed and
trajectory.  As input, the MOT requires the current frame and the positions and
features of actors found in the prior frame.    Output is a set of new positions
and features for all actors detected in the frame.

A separate instance of the second step, PRED, is triggered for each new position
of each actor detected by MOT. PRED predicts the next $q$ positions of
an actor based on the past $p$ positions of that actor.  Thus this
step requires the current position of the actor and the previous $p-1$ positions
tracked by MOT. The output is thus a trajectory composed of $q$ positions. In
our implementation $p=8$ and $q=12$, since the pretrained model employed was
trained with those values.

Each trajectory predicted by PRED triggers a separate instance of the third
step, CD. This step matches the received trajectory with all other trajectories
predicted so far for the same frame, in order to detect collisions.  In addition
to the received trajectory, this step also access all available trajectories
from the same frame. For each frame, after all instances of CD have been
processed, the predicted trajectories of every pair of detected actors will have
been evaluated. The output of each instance of this step is a list of
collisions.

\subsection{RCP Distributed Deployment}
\label{subsec:deploy}

Earlier, we noted that applications such as our RCP example lend themselves to a distributed
deployment.  What we failed to say is that on many platforms, doing so would be a big departure from
what the platform supports.  In fact, today's most common approach for implementing an application
like the RCP is to combine the whole pipeline into a monolithic application, then run it on one
multi-core server, selected by a load balancer.  In order to scale and keep up in real-time, load
balancing tends to randomly spray requests over nodes. Although this approach avoids overheads
associated with transitioning from one step to the next (since all steps happen in the same
process), with complex tasks such as the RCP, it could easily exceed what one server can handle.

Consecutive invocations of the pipeline for frames from the same camera have a
data dependency.  In addition, multiple steps (MOT and PRED) require GPU
accelerators. As a consequence, a frame $k$ must wait until the whole pipeline
for frame $k-1$ finishes processing and make the necessary input data available.
Even frames from other cameras will need GPU resources to be freed.
Furthermore, multiple instances of PRED, one for each actor detected by MOT in
the same frame, can be executed in parallel. However, in the monolithic approach
they would bottleneck, significantly increasing end-to-end latency for that
frame. Finally, certain steps may require more resources than others. In the
monolithic approach, scaling individual steps becomes a challenge. For example,
in our experiments the PRED step presented the heaviest workload, since each
frame had up to 49 actors, and thus this step required more resources.

Thus a job like RCP lends itself to multi-server parallelism: now the scheduler
can place different steps on different servers, taking care not to overload any single server.
However, latency will still be sensitive to data movement
overheads related to the data access patterns of each step. 
We further discuss such overheads next.

\subsection{RCP Data Access Patterns and Overheads}
\label{subsec:issues}

When we deploy an AI pipeline on multiple servers, load balancing will often run each new task on a
server picked at random among lightly loaded machines. Yet because AI models and data dependencies
can be very large, the new task then pauses to fetch data over the network. In real-time
applications such as the RCP, standard solutions such as caching are of limited value due to data
freshness. We discuss next the extent of this effect, and how developer-provided
knowledge can help the platform.

In the MOT step, each request requires a fresh and potentially big data object -- features and
positions of actors in the previous frame (up to 10MB in our experiments).  The nature of the input
(a live stream) implies that these objects will only be used once, hence caching would be
ineffective, since cached objects will never be reused. One way to minimize this overhead would be to send the object to the node where the
{\em next} request will be processed.  However, the MOT step only learns which objects it requires
at runtime, after receiving the request. 

PRED requires small but very fresh objects. Actor positions can be re-used only up to 7 times, and
there are many PRED instances for each frame (one for each actor detected). Thus the overhead from cache misses can add up and
become non-negligible in a time-pressured scenario, as such instances are placed on different servers by the load balancer.
This overhead grows as more nodes are added to
scale out and cache misses are more common. Furthermore, which objects will be
necessary are only known at runtime, after MOT is finished. CD follows a similar rationale: the same
trajectory is accessed many times in a short time (the same frame), and the actual workload is not
predictable.

Despite the freshness and the unpredictability of data access, the developer of the RCP application
knows the mapping between future requests and objects stored previously. A platform can benefit from
this knowledge and collocate correlated objects and requests proactively, slashing data movement
overheads. We achieve that with the \afgrouping mechanism, proposed next. Later, we show how
significant the overheads discussed above can be in the experiments reported in
Sections~\ref{sec:cascade} and~\ref{sec:azure}.

\section{Affinity Grouping}\label{sec:affinity}

In this section, we start 
with an overview of how data and tasks are processed in modern SP and AI serving platforms. We then discuss
requirements for the mechanism arising from the challenges identified in
Section~\ref{sec:motivation}. Finally, we introduce and discuss the \afgrouping mechanism.

\subsection{Execution and Data Flow Model}
\label{subsec:model}

We now describe a general execution and data flow model for a typical platform that hosts
applications such as the RCP.  Our proposals are related to the location at which data is
stored/retrieved, and computational tasks are executed. A location refers to a network endpoint
(e.g. a server node) that holds resources such as CPUs, GPUs, hard disks, and memory. A platform
consists of multiple of such resources distributed across an edge/cloud datacenter network. 

Platforms can be divided in two main subsystems that constitute the {\em platform-level runtime
engine}: storage and compute. As an application runs, data objects are stored and retrieved (e.g.
actors positions in RCP), and computational tasks are initiated (e.g. a step of the RCP pipeline).
In both cases, the corresponding subsystem needs to make placement decisions.  The storage subsystem
contains a data scheduling component responsible for deciding on the location a data object is to be
stored at or retrieved from. Similarly, the compute subsystem has a task scheduling component
that decides upon the location at which a computational task is started. Data objects and/or
computational tasks are said to be {\em collocated} when the corresponding scheduling components
place them at the same location.

In addition, a platform also provides developers with an ap\-pli\-ca\-tion-level API, used to 
interact with the plat\-form-level runtime engine.  
Garbage collection, i.e. how data
objects are cleaned from storage when they are not needed (e.g. actor positions in RCP are not
needed any longer after some seconds), is orthogonal to this work and thus out of scope.


\subsection{Requirements}
\label{subsec:req}

Introduction of feedback by the application developer affects two aspects of a platform: the application-level API, and the plat\-form-level runtime engine. The
API provides developers with a way to encode
ap\-pli\-ca\-tion-specific knowledge about correlations, while the plat\-form-level runtime engine makes
use of the knowledge provided by developers to improve data/computation
collocation, scheduling, prefetching, among other benefits discussed in
Section~\ref{subsec:benefits}.

Ideally, the method of specifying correlations through the application-level API exposed by the platform should be {\em deploy\-ment-agnostic}: the developer logic should not be intertangled with specific deployment details.
For example, the only way to be certain that a computation in Apache Spark Resilient Distributed Datasets (RDDs)~\cite{rdd} will be collocated with data is to manually code a routing mechanism, using internal RDDs APIs to determine which servers have copies of which data items. A similarly complicated proprietary approach is required to ensure collocation in Microsoft Azure, as we show in Section~\ref{sec:azure}. While RDDs offers a way to specify a list of {\em location preferences} when storing data, these are only treated as hints, thus not providing a guarantee.


The method of specifying correlations should also be {\em expressive} enough to allow the platform to capture the data and
computation correlations at runtime. For example, a single computation may require several data
objects, and a single data object may be required by multiple computations. 
A common approach that we reject as inflexible assumes that when coding an application,
developers can already anticipate the data objects required by each computational stage and
statically declare
these dependencies (e.g. using a declarative language or SQL)~\cite{db4ai2,declarative,edgeserve}. The platform then can parse the specification and create
its own internal representation to keep track of which data objects are required by which
computations. However, as seen in the RCP application, an AI model might dynamically fetch large
objects based on analysis of its inputs.  Such model would have an unspecifiable yet
latency-critical data dependency.

Mechanisms such as location preferences in RDD~\cite{rdd}, the {\em \partkey} in Cosmos
DB~\cite{cosmos}, and the {\em \hashtag} in Redis~\cite{redisHashtag} all facilitate collocation of 
correlated data.  For example, positions of actors in the same image could be collocated in our RCP
application. The cloud runtime environment then introduces further mechanisms intended to collocate
stages of a pipelined computation (for example, MOT inference
requests associated with the same RCP video stream) so that a single node handles them.
Other examples of collocating computation include Apache
Spark Streaming {\em partitions} and Apache Storm {\em stream
grouping}~\cite{stormGrouping}. The problem here is that the data collocation features are decoupled from
the computation collocation options.  Developers
need a {\em unified} mechanism that spans the full deployment and covers both storage placement choices and
computational placement decisions. 
Furthermore, it is important that the specification method is {\em easy-to-use} for application developers, i.e. only minimal changes to the application code should be necessary. Modern APIs for developing AI-based applications generally satisfy this requirement.

Finally, the mechanism should be {\em lightweight}: the engine should support specified collocations efficiently at runtime, i.e.  the overhead introduced by the implementation within the engine
should be negligible and remain constant as the platform scales out. Mechanisms such as Redis
\hashtags and Azure \partkeys are potentially costly: the mapping between object keys and labels
must be synchronized across the whole infrastructure, which can put a costly distributed operation
such as a database update on the critical path.

Each of the requirements described above
is not particularly challenging by itself. The challenge comes when trying to satisfy all
requirements together. For instance, the requirement of expressiveness may be in conflict with ease-of-use:
providing more information may also increase the complexity of the API for developers. Similarly,
designing a unified mechanism may also increase the complexity
for developers, and also for the platform to keep track and synchronize correlation information
across the infrastructure. Moreover, a deployment-agnostic mechanism may not be expressive enough,
since deployment details have to be abstracted away. 

\subsection{Affinity Grouping Mechanism}
\label{subsec:mechanism}

Our goals and requirements create a complex problem, particularly in
modern cloud architectures where one often finds completely independent
subsystems specialized for different roles, each with its own design.  In 
particular, storage and computation are generally treated as distinct subsystems,
despite the fact that many compute nodes have substantial caches that can be
viewed as a component of the storage layer.  This observation became the launch
point for our \afgrouping mechanism, which groups correlated objects (and hence
is a storage-layer feature), yet exposes the locations of these groups so that
runtime schedulers can leverage object location data in their decisions.

In \afgrouping, each data object and computational task has a label, similar in spirit to \hashtags in Redis or \partkeys in
Azure EHs.  The critical difference is that labels can be computed dynamically at runtime as inputs are classified. For
example, an input image showing a taxi could be given a label matching various category-specific AI
models and data previously saved in the platform.  We call such labels \afkeys: each object and task
has both a unique name (for example, a file pathname or a K/V key) and an \afkey, which would not be
unique.  Such an approach allows the platform to derive correlations between data and computation: any data objects and a task with the same \afkey will be regarded as requiring correlation. 



The core of the proposed mechanism is a function $f(d)$, which maps a descriptor $d$ to
an \afkey.
A descriptor contains metadata about a data object (to be stored or retrieved) or a
computational task (to be initiated).
Affinity keys are labels that can be implemented e.g. as strings.  The
application-level API allows developers to provide the \afgrouping function $f$
to the platform. When a placement decision for a data object or a computational task needs to be
made, the platform then can apply function $f$: the function 
extracts information from the given descriptor to generate an \afkey.
Application-specific knowledge is thus entirely encapsulated in $f$.  Note that
$f$ will be available throughout the distributed service, and must return the
same result for a given descriptor no matter where it is invoked.  Moreover, it is
often invoked on critical paths, where blocking would be problematic.  For
example, in Section~\ref{sec:cascade} we discuss an \afgrouping function that
uses regular expressions.

The platform-level runtime engine must guarantee that the location where data objects are
stored and/or cached, and computational tasks are placed, depends on \afkeys.  In the RCP
application, for example, $f$ can map data objects of past positions of an actor $a$ to a label
``actor\_a''. The computation task, corresponding to the PRED step, predicting the future trajectory of
actor $a$ can also be mapped to the same label, letting the platform know at runtime that the
corresponding objects and the computation are correlated. 

Regarding the requirements defined above, the
proposed mechanism is deployment-agnostic, since function $f$ maps descriptors to \afkeys according
only to application-specific knowledge. The platform is then responsible to decide how to handle
objects and tasks from each label according to the deployment. The mechanism is unified, since function $f$ applies to
placement decisions on both storage and compute subsystems.
Tasks and objects are linked through
labels, instead of an explicit static list of input objects or query. Developers have a great deal
of flexibility when expressing data access patterns of their application (i.e. expressive).  Turning
to the efficiency requirement, only function $f$ itself must be available on all
components of a platform: there is no associated replicated state. Compliance with the ease-of-use
requirement, however, depends on the specific implementation. Section~\ref{sec:cascade} describes an
implementation in which developers only need to provide a regular expression.

\subsection{Potential Benefits}
\label{subsec:benefits}

\begin{figure}
  \centering
  \includegraphics[width=0.7\linewidth]{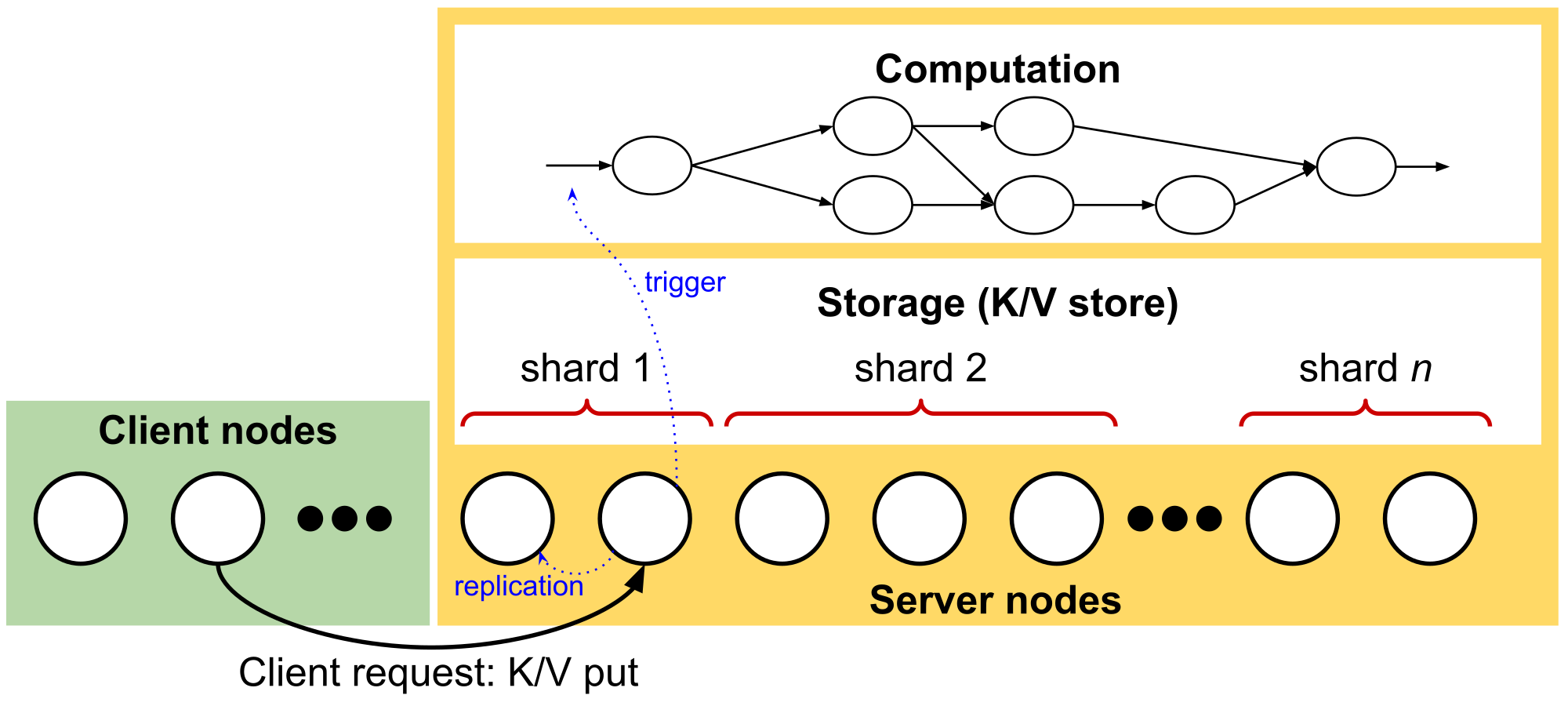}
  \caption{Cascade architecture.}
  \label{fig:cascade_arch}
  \vspace{-6mm}
\end{figure}

A platform-level runtime engine can take advantage of \afkeys in many
different ways. 

\itemParagraph{Proactive collocation:}
It is possible to proactively collocate correlated data and computation by
factoring \afkeys into the caching infrastructure, as well as into placement
decisions of both computations on servers and objects
within a scalable storage infrastructure.
For example, in the RCP application, all positions from the same
actor (output from the MOT step) can be stored in the same physical node where
the PRED computation for that actor will take place when/if it arrives. In this
example, the \afkey for all objects and tasks should
be the same (the identifier of the actor, for example). Collocation avoids the
extra overhead of fetching input data from remote processes.

\itemParagraph{Prefetching:}
The knowledge encoded into \afkeys can also enable proactive decisions in anticipation of a future
need. For example, in a multistage job a scheduler launching the first stage may decide to prefetch
objects that will be needed by a downstream stage, based on the \afkey of the objects and the
corresponding stage, in anticipation of the computational task reaching the corresponding nodes.


\itemParagraph{Consistency:}
Many edge AI applications are sensitive to data sequencing.  
Similarly to \partkeys in Azure EH, \afkeys can be used to guarantee 
consistency. Objects and tasks with the same \afkey may have to be handled
sequentially and in order. Additionally, a platform may allow objects sharing the same \afkey to be updated atomically, in a single action: this was easily achieved in our prototype, since objects with the same \afkey are stored in a same shard, as described later in Section~\ref{sec:cascade}.

\itemParagraph{Parallelism:}
Objects and tasks with different \afkeys have no mutual dependency and may thus
be handled in parallel. Furthermore, the \afgrouping mechanism enables a finer
grain of parallelism control compared, for example, with a mechanism such as
Azure EHs. In EHs, requests with different \partkeys may be placed in the same
partition, causing the requests to be handled sequentially.

\section{Evaluation on Local Cluster}\label{sec:cascade}

In this section, we describe an implementation of the \afgrouping mechanism in
Cascade~\cite{cascade_arxiv}:  a state-of-the-art stream processing
platform targeted at latency-sensitive AI workflows. Cascade was selected
because it offered the highest baseline performance among open source platforms
amenable to our methodology, enabling us to ask whether even lower latency and
higher throughput might be feasible using an \afgrouping methodology.

The goal of this section is to show the benefit of \afgrouping on the end-to-end
(E2E) latency of the RCP application. We evaluate the E2E latency of the
application with different workloads, while also scaling out the Cascade
deployment. Section~\ref{sec:azure} provides a deeper analysis on the overheads
in each step of the pipeline. 

\subsection{RCP Application Implementation}
\label{subsec:app_impl}

Our experiment used the Stanford Drone Dataset (SDD)~\cite{sdd}, which contains
aerial videos showing campus intersections.  The videos include several types of
actors (e.g. pedestrians, cars, cyclists).  Data source ``clients'' simulate
cameras by streaming these videos to the pipeline. The MOT and PRED steps are
based on off-the-shelf AI models, and we used the source code and pre-trained
models provided by the authors as much as possible: most of our modifications
focused on data ingress/egress. All three steps are implemented in Python using
the PyTorch library.

To implement the MOT step, a multi-object
tracker~\cite{yolov5-strongsort-osnet-2022} that employs YOLO5 for actor
detection was used.  StrongSORT~\cite{strongsort} and OSNet~\cite{osnet} are
used for re-identification and trajectory tracking.  We selected three videos
from the SDD for training (\little, \hyang, and \gates).  Data is uncompressed,
hence each frame is approximately 8MB in size (compressing and then
decompressing frames is slower than just transferring them uncompressed).  State
data containing positions and features of actors detected in a frame (required
to re-identify actors in the following frame) ranges from a few KBs to around
10MB depending on the number of actors in the frame.  Frames from the same video
must be processed sequentially, but frames from different videos can be
processed in parallel by multiple MOT instances.

The PRED step employs a trajectory prediction model, YNet~\cite{ynet} pretrained
on the SDD data by the authors~\cite{ynetCode}. PRED requires eight consecutive
positions for each actor and makes no prediction if fewer than eight are
available. Position objects are small: 10s of bytes.  Instances of PRED for the
same actor must be processed sequentially. Instances triggered for different
actors may be executed in parallel. Notice that the PRED workload depends not
only on how many clients are streaming frames, but also on how many actors are
detected in each frame by MOT.

CD consists of a simple algorithm that performs a linear interpolation
on the predicted trajectories of identified actors, outputting a warning if any
pair crosses.  Instances of this step corresponding to the same frame and client must be processed sequentially to ensure that all pairs of
trajectories will be matched with each other. Different frames (even if from the
same client) may be processed in parallel.

\subsection{Cascade: System Architecture}
\label{subsec:cascade_model}

Cascade~\cite{cascade_arxiv} is a full-stack platform for high-speed stream processing that
prioritizes low latency by hosting data and compute, avoiding copying and
locking on critical data paths and leveraging acceleration technologies such as
RDMA and DPDK. 
It consists of a set of nodes (clients and servers) interconnected in a complete
network, typically in an edge/cloud datacenter.  Client nodes may issue requests
to any server node. Server nodes implement two subsystems: storage and
computation.  Each subsystem is further described next.
Figure~\ref{fig:cascade_arch} shows an overview of the system architecture.


The storage subsystem implements a sharded Key/Value (K/V) object store.
Server nodes are logically grouped into disjoint {\em shards}.
Prior to our work, each object key was hashed to determine the {\em home}
shard for the K/V pair.  Cascade supports multiples levels of persistence, but
our work considered only the two in-memory mode: {\em trigger} requests, which
cause a task to run but leave no data behind; and {\em volatile storage} requests, 
which replicate an object that will then be retained in memory by all members
of the home shard.

Cascade's {\em User-Defined Logic} (UDL) framework is responsible for executing
computational tasks.  The code for each task is supplied by the developer as a
container or in a Dynamic-Link Library (DLL). In either case, a task is associated with a {\em
key prefix}.  On each Cascade server, an upcall will occur if that server
receives a K/V pair with a matching key.  For example, if a task is registered
using the prefix ``/RCP/taxis'', then an update to a K/V pair with key
``/RCP/taxis/1234'' would trigger that task at whichever node the put was sent to.
Each task can also issue another put, enabling pipelined tasks in which each stage initiates the next stage.
In the context of this work, a potential overhead that we want to minimize
occurs when a task reads an object that is not homed on the node
where it is executing. The object then must either be fetched from cache or over
the network.  Even with high-speed networking primitives supported by RDMA or
DPDK, network transfers of large objects are costly.

Cascade supports a form of resource partitioning called an {\em object pool}.
Rather than treating all Cascade nodes as a single
sharded service, the nodes are instead organized into groups.  A single server
node can belong to multiple pools.  Cascade allows the developer to configure
each pool with its own properties, such as the shard size to use, and the degree
of data persistence the pool will offer.
For example, a volatile pool
can be pinned to host memory, GPU memory, etc.    Pools are identified by
pathname prefixes: if /x/y is the prefix of a volatile pool holding GPU memory
objects, /x/y/z could name a tensor residing within that GPU memory.

\subsection{Affinity Grouping on Cascade}
\label{subsec:cascade_imp}

As noted earlier, requests in Cascade consist of
{\em trigger}, {\em put} or {\em get} operations. The first two are
parameterized by the key and value (uninterpreted byte vector) of the object
being transmitted/saved, whereas get takes a key and returns the corresponding
object.  We implemented function $f$ as a regular expression, which is matched
against request keys (the descriptor).
The \afkey is a substring of the request key,
composed of the characters that matched the regular expression. We employed the
Hyperscan library~\cite{hyperscan} for matching regular expressions, which
introduced a negligible overhead to Cascade's critical path: according to
microbenchmarks, matching the regular expressions employed in the RCP
application was under 300 microseconds on average.

To implement the application-level API of the \afgrouping mechanism, we extended
the Cascade client API. The extension allows developers to register  regular
expressions with a specified \objectpool. Listing~\ref{list:code_objectpool} shows
a code excerpt (in Python) with an example of how \objectpools are created, with
and without employing \afgrouping.
The only modification made to the Cascade API was the addition of the optional argument \texttt{affinity\_set\_regex} to the method that creates an object pool (line 12).
Any put or get operation assigns an \afkey to the corresponding object by matching the object key against the regular expression registered in the corresponding object pool (line 18).

\definecolor{cadetblue}{rgb}{0.37, 0.62, 0.63}
\definecolor{darkolivegreen}{rgb}{0.33, 0.42, 0.18}

\begin{lstlisting}[
    caption={Creating pools with/without \afgrouping.},label=list:code_objectpool,captionpos=t,float,abovecaptionskip=-\medskipamount,
    linewidth=0.9\linewidth,
    language=Python,
%    frame=single,
    numbers=right,
    basicstyle=\tiny,
%    backgroundcolor=\color{white},
    keywordstyle=\ttfamily\color{darkolivegreen},
    identifierstyle=\ttfamily\color{cadetblue}\bfseries,
    commentstyle=\color{brown},
    stringstyle=\ttfamily,
    showstringspaces=true]
capi = ServiceClientAPI() # Cascade client API
subgroup_type = "VolatileCascadeStoreWithStringKey"
subgroup_index = 0

# creating object pool without affinity grouping
capi.create_object_pool("/no_grouping",
                        subgroup_type,subgroup_index)

# creating object pool with affinity grouping
capi.create_object_pool("/grouping",
                        subgroup_type,subgroup_index,
                        affinity_set_regex="_[0-9]+")

# putting in the object pool that is not grouped
capi.put("/no_grouping/example_1",None)

# putting in the object pool that is grouped
capi.put("/grouping/example_1",None) # affinity key is '_1'
\end{lstlisting}

With respect to the platform-level runtime engine, we made two modifications to
Cascade: (i) the mapping from key to shard within an object pool is based on the
\afkey instead of the request key; and (ii) the \afkey functions and their
matching expressions are registered in all nodes. As described above, Cascade
previously selected the shard that will handle a request by hashing the object
key supplied with the object describing the request. Our modified policy selects
the shard by hashing the \afkey. Jointly, these changes ensure that all objects
with the same \afkey will be stored (and replicated) in the same shard, and
route requests using that same \afkey to this shard.


\subsection{Computing Environment}
\label{subsec:cascade_setup}

Servers in our local cluster are equipped with Mellanox ConnectX-4 VPI NIC cards connected to
a Mellanox SB7700 InfiniBand switch, resulting in a
RDMA-capable 100Gbps network backbone. All servers have their clocks
synchronized using PTP~\cite{ptp}, making timestamps from different servers
comparable with sub-millisecond precision.  Servers have two configurations,
denoted \textbf{A} and \textbf{B} in this work.  Configuration \textbf{A}
consists of two Intel Xeon Gold 6242 CPUs, 192 GB of memory, and an NVIDIA Tesla
T4 GPU. 
Configuration \textbf{B} consists of two Intel Xeon E2690 v0 CPUs and 96 GB of memory, but no GPU. Our experiments employ up to 8 servers with configuration \textbf{A}, and up to 9 with \textbf{B}.

Cascade was deployed with three object pools, one per step. Cascade nodes in
pools responsible for MOT and PRED were deployed on servers
with configuration \textbf{A}, since these steps require GPUs for AI inference.
Clients and other Cascade nodes were deployed on servers 
\textbf{B}. In this section, we report results with a varying
number of shards in each pool. Each shard has always only one node (i.e. a
physical server): increasing
the size of each shard would increase replication, which is not relevant for
evaluating our proposals. Each configuration of shards is called a {\em layout},
and we denote a layout as $x/y/z$, where values $x$, $y$, and $z$ indicate the
number of shards for steps MOT, PRED, and CD, respectively.

Although Cascade includes a scheduler, we configured the system to place objects
and trigger computations purely by \afkey hashing.  This enables us to focus on
the degree to which a purely \afgrouping placement of data and computation can
improve latency.  

\subsection{RCP Deployment on Cascade}
\label{subsec:cascade_depl}

Each step of the RCP application was deployed as a DLL on Cascade's UDL framework. UDLs are loaded
by each Cascade node on startup, along with models, weights, and hyperparameters. All required
objects (MOT state data, actor positions and trajectory predictions) are stored/retrieved using
Cascade K/V store.

Each client is responsible for a different video, and sends video
frames at a rate of 2.5 FPS (the rate the AI model employed in PRED was trained
at). To send a frame, a client puts an object in the K/V store with a key that
will trigger a MOT task. The MOT task retrieves the state data from the previous
frame, and puts a separate object for each actor position found in the frame.
Each actor position put by the MOT task triggers a separate PRED computation.
Each PRED task first retrieves the past positions of the corresponding actor
from the K/V store, and puts an object with the trajectory prediction. Each
trajectory prediction triggers a separate CD computation, which first retrieves
all trajectory predictions for the same frame available in the K/V store, and
then performs the computation. The result of the collision detection is put in
the K/V store. The E2E latency is the time elapsed between the client sending
the frame, until the result of the last collision detection for that frame is
put in the K/V store.  All tasks cache in memory all objects they retrieve or
create.

We employed two workloads: single clients (\little, \hyang, and
\gates), and three simultaneous clients (\littlehyanggates). Each client sends
700 frames, and we discard measurements from the first 100 frames. Each
experiment is repeated three times. Cascade was completely cleared of objects
and \objectpools before each run of each experiment.

We grouped requests from each step using the \afgrouping mechanism in a different way. For
MOT, grouping was based on the identifier of the client, so that all frames from
the same client always went to the same shard. For PRED, grouping was based on
actor identifiers, and for CD it was based on both the client identifier and the
frame number. Table~\ref{table:keys} shows all the employed \objectpools,
key examples, triggered tasks (if any), the regular expressions used to group
requests (if any), and the resulting \afkeys.

\begin{table}
\centering
\newcolumntype{x}[1]{>{\centering\arraybackslash\hspace{0pt}}m{#1}}
\renewcommand{\arraystretch}{1.2}
\setlength{\tabcolsep}{2pt}

\tiny
\caption{Object pools, example keys, triggered steps, regular
    expressions, and affinity keys.}
\label{table:keys}
\begin{tabular}{c|c|c|c|c}
\textbf{Object Pool}                    & \textbf{Example Key}                                   & \textbf{Step} & \textbf{Regex}                  & \textbf{Affinity Key} \\ \hline
    \texttt{/frames}             & \texttt{/frames/little3\_42}                & MOT                  & \texttt{/{[}a-zA-Z0-9{]}+\_}             & /\texttt{little3\_}            \\
    \texttt{/states}              & \texttt{/states/little3\_42}                 & --                   & \texttt{/{[}a-zA-Z0-9{]}+\_}             & /\texttt{little3\_}            \\
    \texttt{/positions}     & \texttt{/positions/little3\_7\_42}     & PRED                 & \texttt{/{[}a-zA-Z0-9{]}+\_{[}0-9{]}+\_} & /\texttt{little3\_7\_}         \\
    \texttt{/predictions} & \texttt{/predictions/little3\_42\_7} & CD                   & \texttt{/{[}a-zA-Z0-9{]}+\_{[}0-9{]}+\_} & /\texttt{little3\_42\_}        \\
    \texttt{/cd}                    & \texttt{/cd/little3\_42\_7\_5}                 & --                   & --                              & --                   
\end{tabular}
  \vspace{-3mm}
\end{table}

Results reported in this section compare two different placement strategies,
random placement and \afgrouping. In the random placement strategy, the RCP
application is executed without using \afgrouping, i.e.  objects and
computations are placed randomly by hashing requests keys -- the standard
Cascade behavior. In \afgrouping, we group requests as described
above. Hash-based pseudo-random mapping from key to shard is standard in
key-value stores. Our discovery is that by combining \afgrouping with
randomization on the affinity key, we get the best of both worlds: excellent
load-balancing and scaling with sharply higher cache hit rates, hence less data
movement.

\subsection{Results}
\label{subsec:cascade_results}

\begin{figure}
    \centering
    \includegraphics[width=0.85\linewidth]{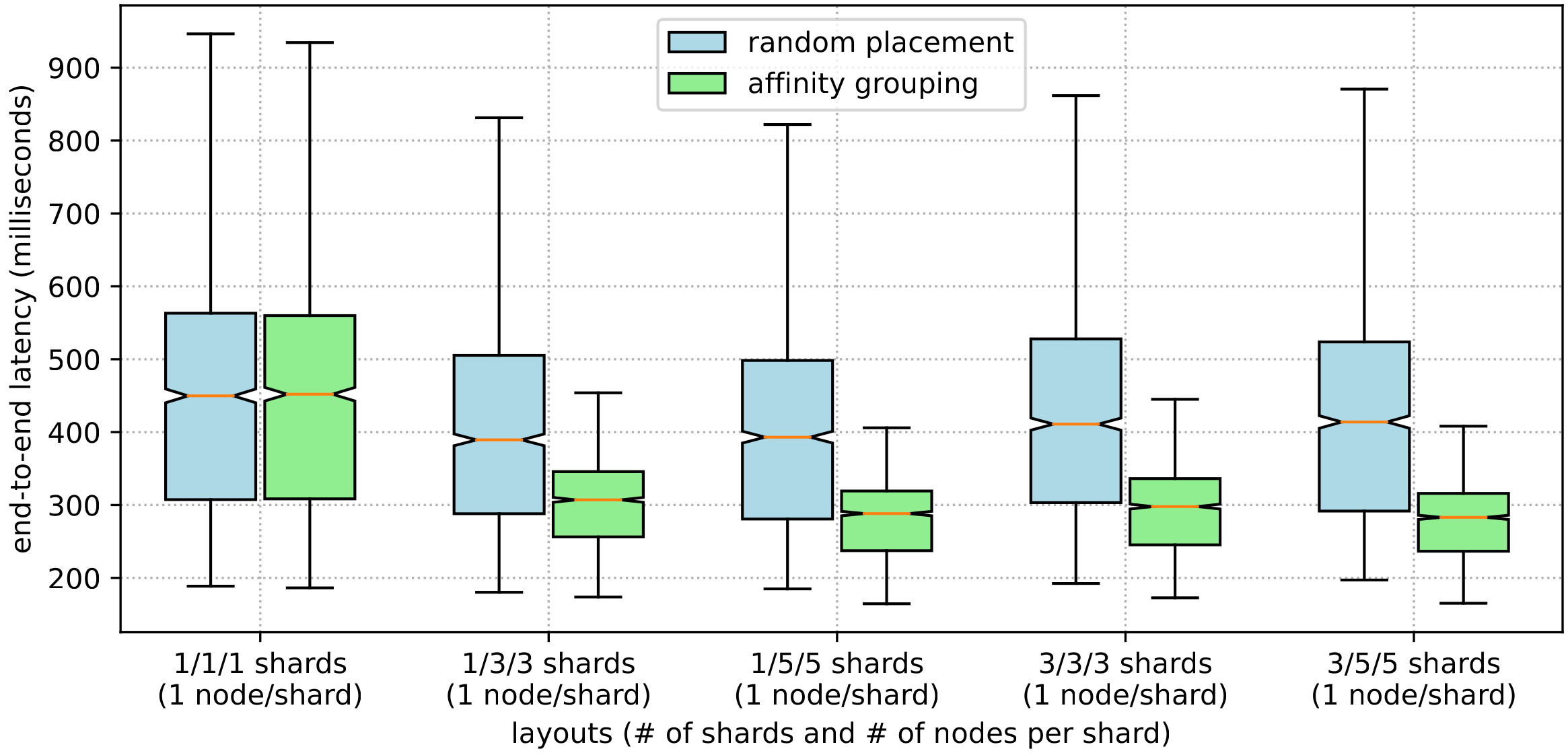}
    \caption{E2E latency for \gates on Cascade.}
    \label{fig:cascade_1cam_gates3}
  \vspace{-2mm}
\end{figure}

\begin{figure}
    \centering
    \includegraphics[width=0.85\linewidth]{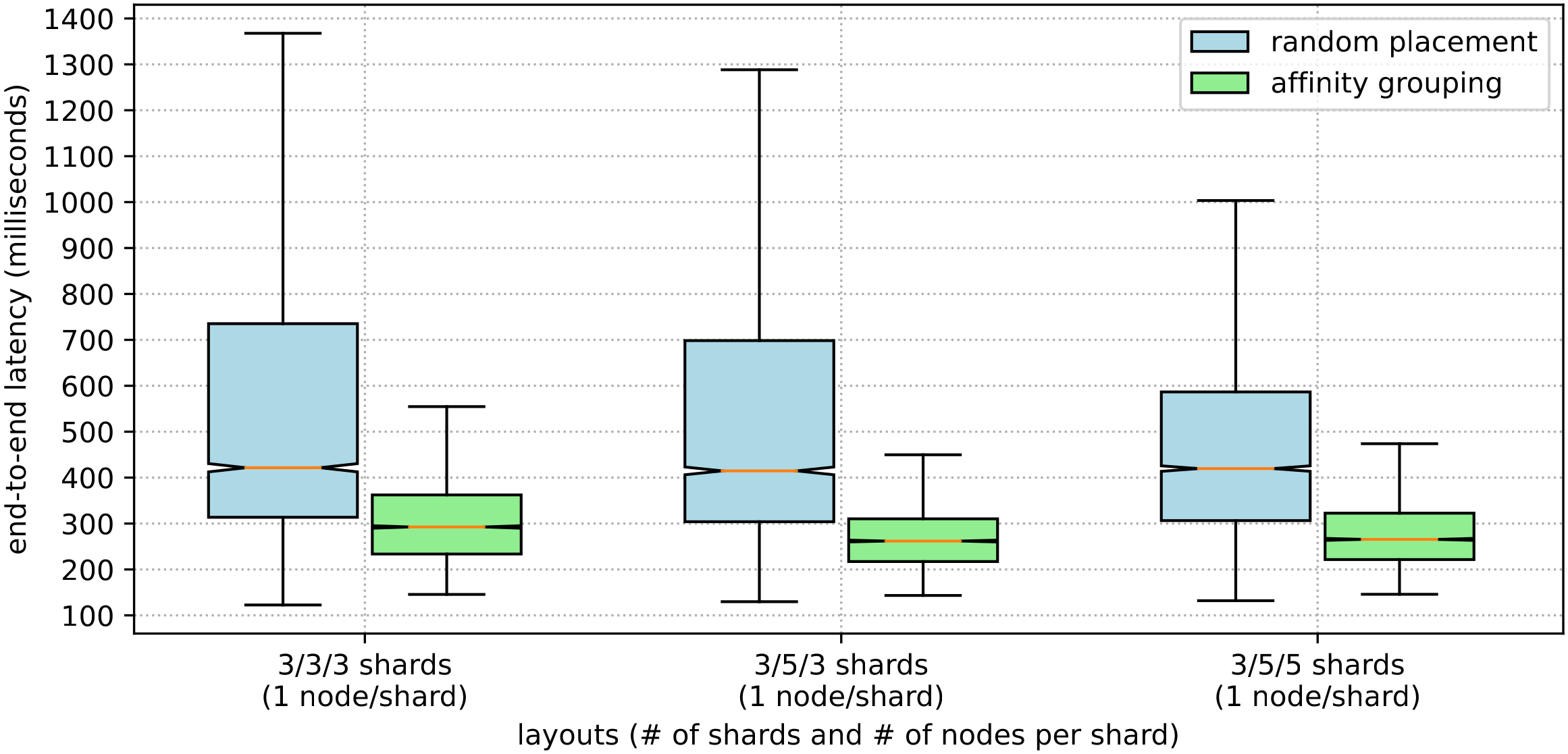}
    \caption{E2E latency for three clients on Cascade.}
    \Description{E2E latency for three clients on Cascade.}
    \label{fig:cascade_3cam}
  \vspace{-3mm}
\end{figure}

We now report results for just one client, varying the number
of shards for MOT (1, 3), PRED (1, 3, 5) and CD (1, 3, 5).
Figure~\ref{fig:cascade_1cam_gates3} shows a box plot of the E2E latency (in
milliseconds) for different layouts and placement strategies. We show only client \gates, 
results for other clients followed the same pattern. Both median and 75th percentile latencies were significantly
reduced by \afgrouping, except for layout 1/1/1: grouping
had no effect since there was only 1 shard per step. Adding shards does not help with the random placement strategy. Although increasing from 1/1/1 to 1/3/3 did help, increasing the number of shards even further results in higher
median and percentile latencies, as fetching overheads increase due to more cache misses.
We now increase the workload. Figure~\ref{fig:cascade_3cam} shows the E2E latency for three
simultaneous clients (\littlehyanggates), employing different layouts and
placement strategies. Latency was significantly lower and more consistent as the
deployment scaled out when employing \afgrouping.

\begin{figure}
    \centering
    \includegraphics[width=0.85\linewidth]{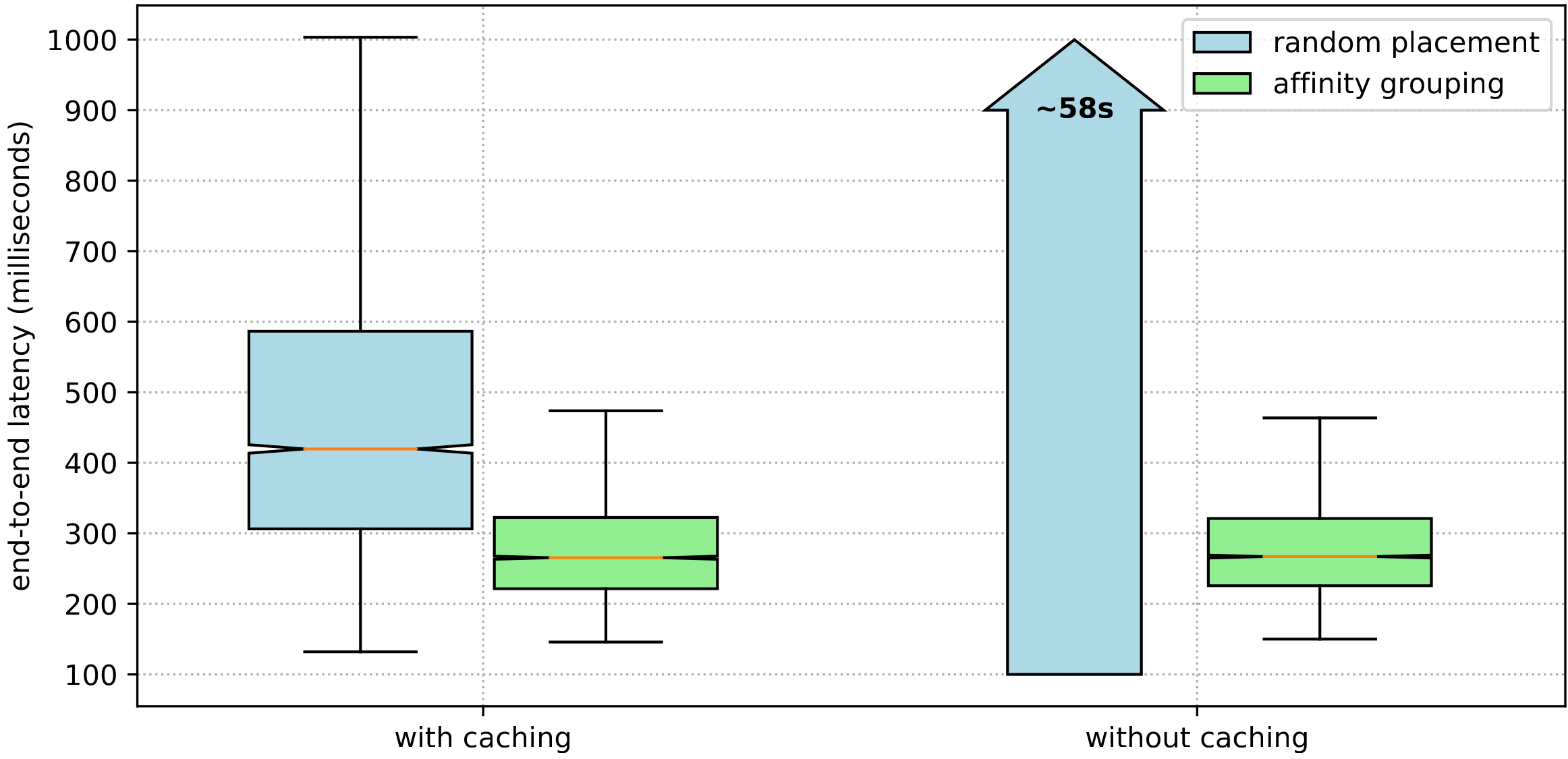}
    \caption{E2E latency with/without caching on Cascade.}
    \label{fig:cascade_nocache}
  \vspace{-3mm}
\end{figure}

To further highlight the benefit of collocation, we disabled object
caching in our application. Each step then always
has to fetch objects from the K/V store. Here, the main difference
between random placement and \afgrouping is that the latter
guarantees that objects being fetched are always stored locally, in the
same Cascade node (and shard) where the task is running. For
random placement, there is no such guarantee.
Figure~\ref{fig:cascade_nocache} shows the results
for three clients and 3/5/5 shards. 
Due to the zero-copy design of Cascade, E2E latency was the
same with or without caching for \afgrouping, since memory copies are
minimized and there is no serialization overhead when an UDL makes a get request
to the same node where it is running: keeping objects in the application memory
or fetching them locally from the K/V store incurs virtually the same cost.
However, for random placement, disabling caching significantly
increased latency and reduced throughput. The median latency was off-scale
at more than 58 seconds, hence the bar is replaced with an arrow.
Throughput was on average 6.7 FPS, thus the pipeline was not able to handle the
7.5 FPS sent by clients collectively.

\begin{figure}
    \centering
    \includegraphics[width=0.85\linewidth]{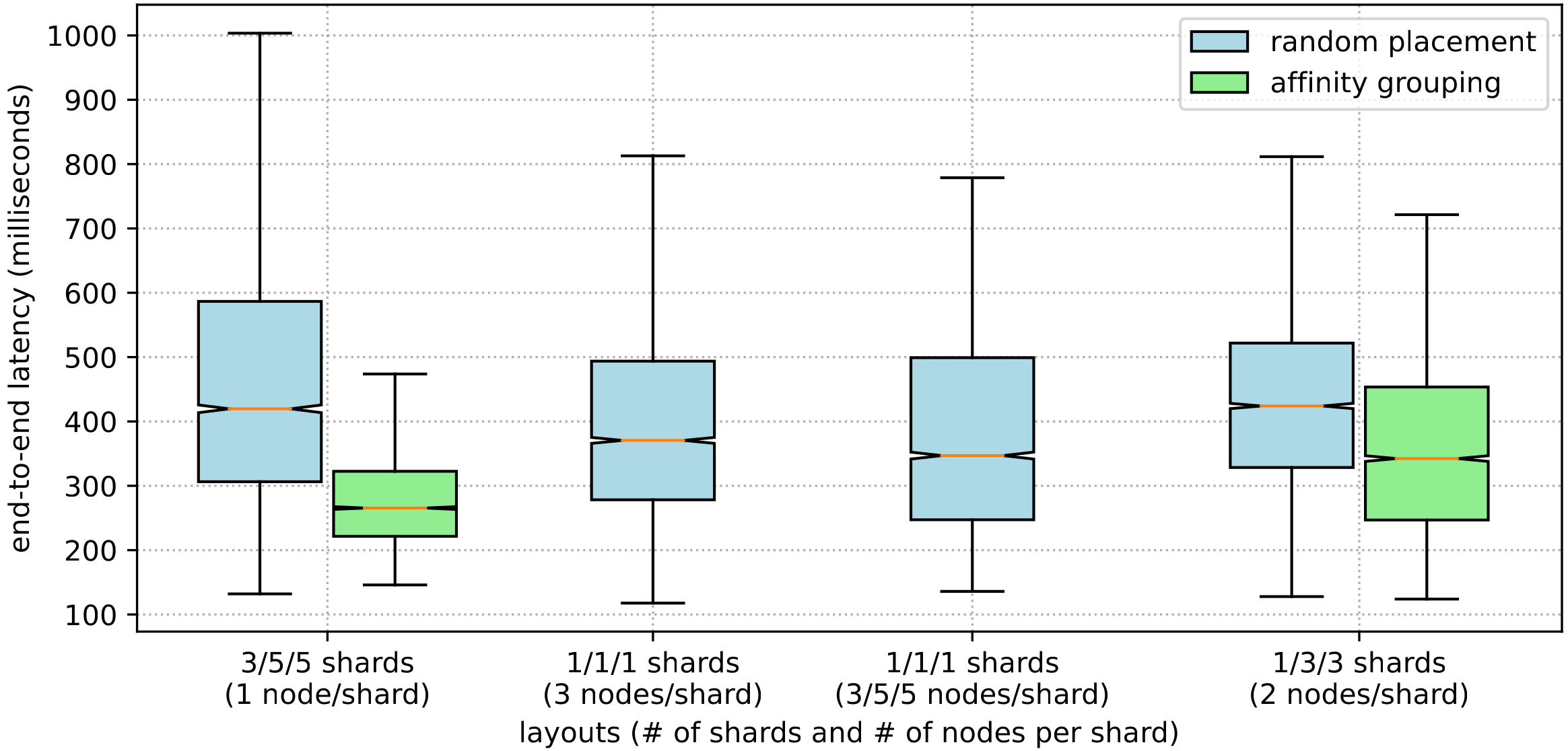}
    \caption{E2E latency with replication on Cascade.}
    \label{fig:cascade_replicated}
	\vspace{-5mm}
\end{figure}

We also evaluated the impact of replication on E2E latency.
Different layouts with more than one node per shard were employed.
In Cascade, when a shard has more than one node, any object
stored in that shard is replicated to all nodes in the shard, before any
task is triggered. As a result, tasks in such a shard
will have local access to the objects in that shard, even if the previous
computation (e.g. state of the previous frame) was performed in another node.
This behavior can be seen as a form of prefetching, since nodes store objects
that will be used in the future before the corresponding request arrives.
However it incurs extra latency since the corresponding computation is only
triggered after the replication is completed, while a more sophisticated
prefetching feature would not prevent the computation to start while 
prefetching is done in other nodes. Figure~\ref{fig:cascade_replicated} shows results for different layouts. As a reference, 
the first group of bars show the latency for
3/5/5 shards, one node per shard (no replication). The next two bars
show results for 1/1/1 shards, with different number of nodes per shard (3
and 3/5/5), so there is no difference between random placement and \afgrouping.
The last layout has 1/3/3 shards, two nodes per shard, configuring a
compromise between many shards with just one node each and a single shard
with many nodes. Results show that replication reduced latency compared to the
baseline (first bar). However, employing \afgrouping and multiple shards still
results in better latency. We argue that these results show the potential of
further investigating the integration of \afgrouping with prefetching and
scheduling.

\subsection{Insights}
\label{subsec:cascade_dis}

Affinity grouping significantly reduces the latency of the RCP application on
Cascade, compared to other placement options. Although the sharding policy in Cascade makes it scalable by design, results show that \afgrouping
improved Cascade scalability even further compared to the standard random placement, where adding nodes sometimes
degraded performance due to increased cache misses. It is important to note that the scalability of \afgrouping is orthogonal to the scalability of the system as a whole. The scalability of the \afgrouping mechanism itself is discussed in Section~\ref{sec:affinity}.

Changes to the application were primarily to register
regular expressions (Table~\ref{table:keys} during initialization, as
illustrated in Listing~\ref{list:code_objectpool}).   Support for \afgrouping
required minor changes to Cascade itself, yet enabled the system to use \afkeys
to place objects and route requests. 
Our results on replication indicate that there is potential to
improve latency even further by integrating \afgrouping with prefetching.

\section{Evaluation on Azure Cloud}\label{sec:azure}

A natural question to ask is whether our claim that existing SP and AI serving
platforms are inadequate is correct.  To address this, we now describe a second
deployment of RCP on Microsoft Azure Cloud. Granted, Azure is a public cloud
whereas Cascade primarily targets a private cluster, but we believe this is a
fair comparison because it genuinely represents a state-of-the-art alternative.

The goal of the section is to show evidence that: (i) data access patterns of
applications such as RCP indeed result in extra fetching overheads; and (ii)
although it is possible to overcome these overheads, it comes at the cost of a
high coupling between application and deployment, since there is no unified
mechanism available in modern platforms such as the proposed \afgrouping.

\subsection{Deployment on Microsoft Azure Cloud}
\label{subsec:azure_depl}

\begin{figure}
  \centering
  \includegraphics[width=0.9\linewidth]{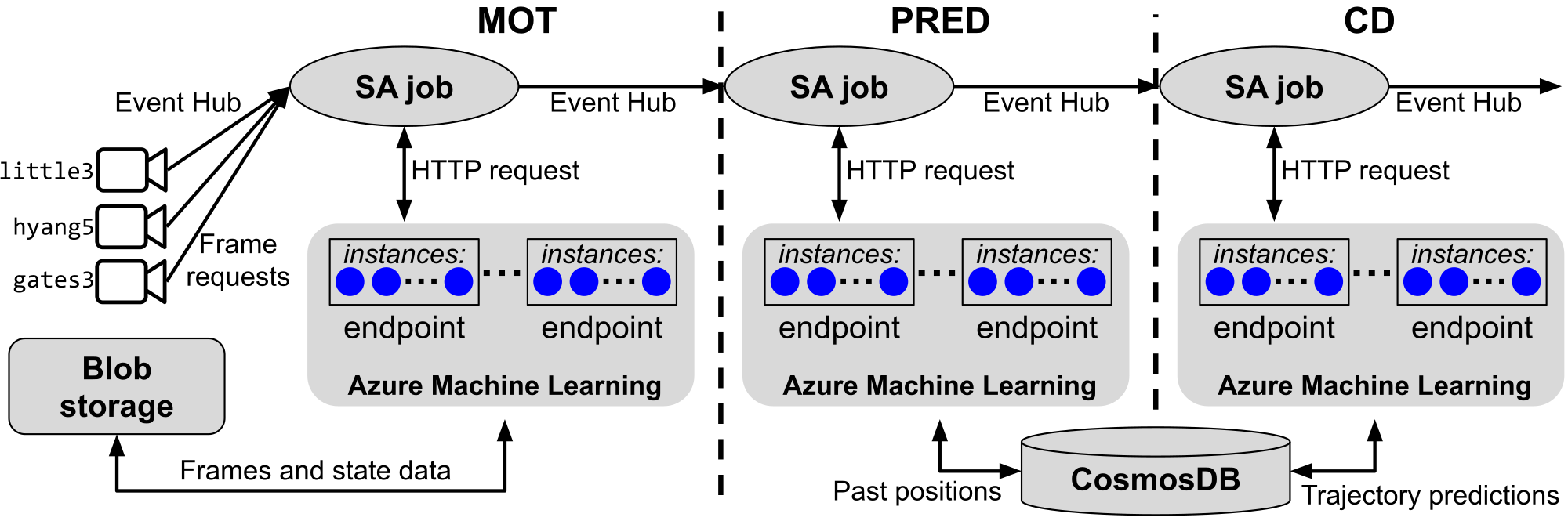}
  \caption{RCP deployment on Azure.}
  \label{fig:app_azure}
  \vspace{-5mm}
\end{figure}

The RCP pipeline was implemented on Stream Analytics (SA) -- Azure's SP
platform. We reused the source-code of each of the 3 steps (MOT, PRED, and CD),
with few modifications. Each of the AI models was first configured on Azure
Machine Learning (AML) as real-time endpoints, using a web-services interface.
AML is elastic: depending on load, endpoints can be backed by worker instance
pools of varying size~\cite{asaandml}. Pipeline stages were connected using
Azure Event Hubs (EHs)~\cite{kettner2022iot}. Clients were deployed in a virtual
machine. Frames from all three videos were stored in Azure Blob
Storage~\cite{calder2011windows}, which was also used to store MOT state data.
Positions of actors and trajectory predictions are stored using Cosmos
DB~\cite{cosmos}. The resulting architecture is shown in
Figure~\ref{fig:app_azure}, and is compliant with recommendations from the cloud
vendor. The individual AI tasks were properly configured with ample resources
(similar to the resources employed previously in the Cascade experiments).

Clients stream frames to the first SA job in the pipeline at 2.5
FPS, through an EH. The SA job invokes an MOT endpoint: one instance in
the endpoint is selected by a load balancer to serve the request.  The instance
fetches the video frame from Blob storage and performs the inference (we assume
the client has already uploaded the frame). The state from the previous frame is
also fetched from Blob storage if it is not already cached in the instance's
memory. Results are sent through another EH to the next SA job responsible for
invoking PRED endpoints for actor positions. For each invocation, the selected
instance of the invoked PRED endpoint stores the new position on Cosmos DB, and
also fetches the past positions of the corresponding actor that are not already
cached in memory. The CD step is performed in a similar fashion. The final
output of the pipeline is sent back to the client running on the virtual
machine. End-to-end latency of a frame is the time elapsed from when the frame
request was first sent by the client, until the last output from the CD step for
that frame was received by the client. 

All instances in AML endpoints maintain MOT states, positions and trajectory
predictions cached in memory for later reuse, thus avoiding fetching from Blob
storage or Cosmos DB as much as possible. In the experiments reported in this
section, we employ, for each step in the pipeline, a varying number of endpoints
and instances per endpoint, as indicated. Instances in MOT and PRED endpoints
were of type $\textit{Standard\_NC4as\_T4\_v3}$ (equipped with NVIDIA T4 GPUs),
while CD instances were of type $\textit{Standard\_DS3\_v2}$. 

We employed as many partitions in EHs and streaming units in SA as supported by
the endpoints (2*number of instances),  maximizing
parallelism~\cite{azureParallel}. The \partkey for MOT requests was the video
name, ensuring that frames originating in any single client were processed
sequentially. Similarly, the \partkey for PRED was the actor identifier, and for
CD the frame number. We set the batch size when invoking AML endpoints to 1, a
configuration intended to minimize per-frame latency.

Our experiments stream 700 frames, but we discard measurements the first 100 to
give the framework time to warmup.  Each experiment was repeated three times.
To vary the workload (the number of simultaneous clients streams), we first
tested with a single video at a time (\little, \hyang, and \gates), then with
two concurrent video streams (\littlehyang), and finally with three
(\littlehyanggates).  

\begin{figure}
  \centering
  \includegraphics[width=0.85\linewidth]{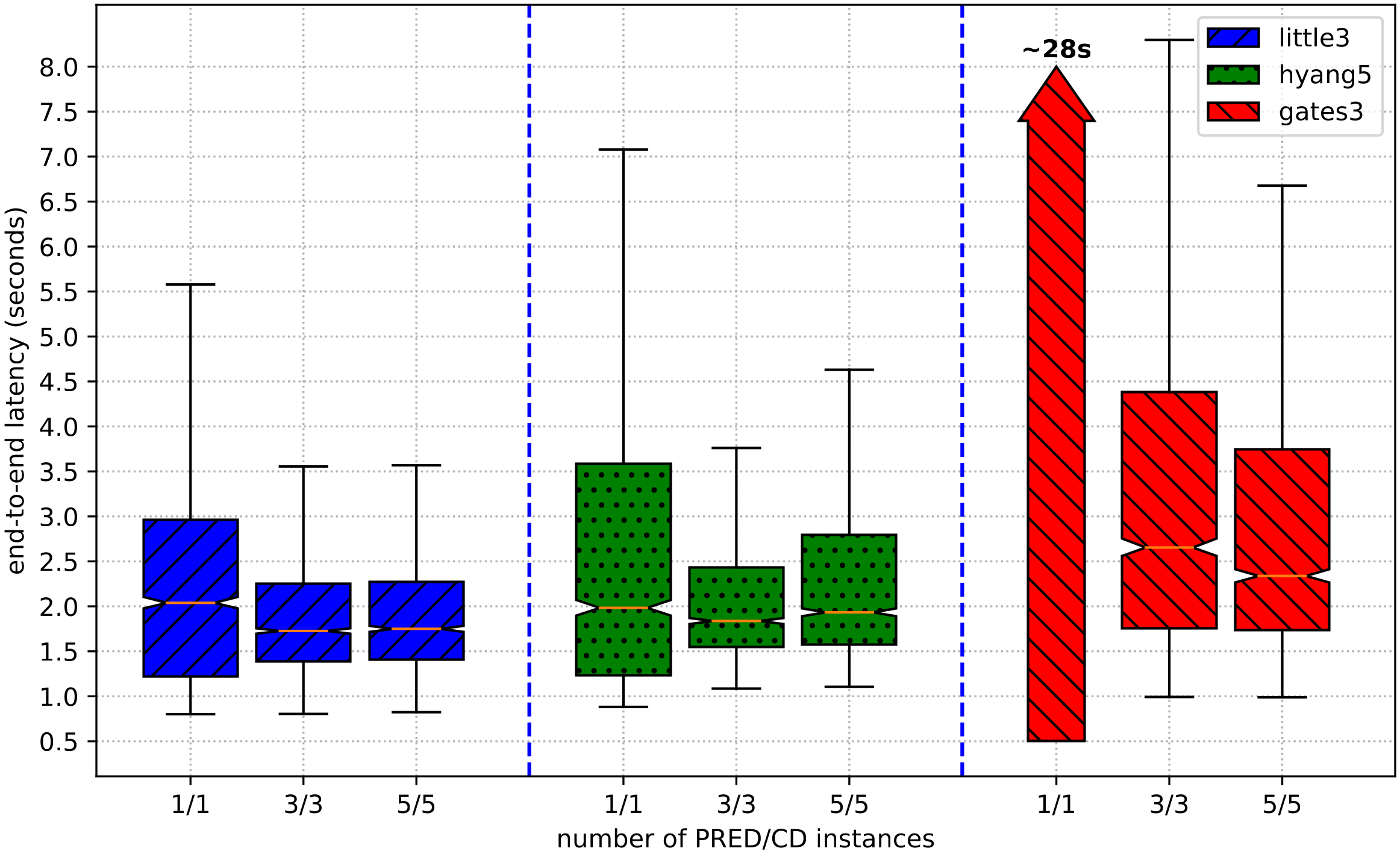}
  \caption{E2E latency for individual clients.}
  \label{fig:e2e_1cam}
  \vspace{-5mm}
\end{figure}

\begin{figure}[h]
  \centering
  \includegraphics[width=0.85\linewidth]{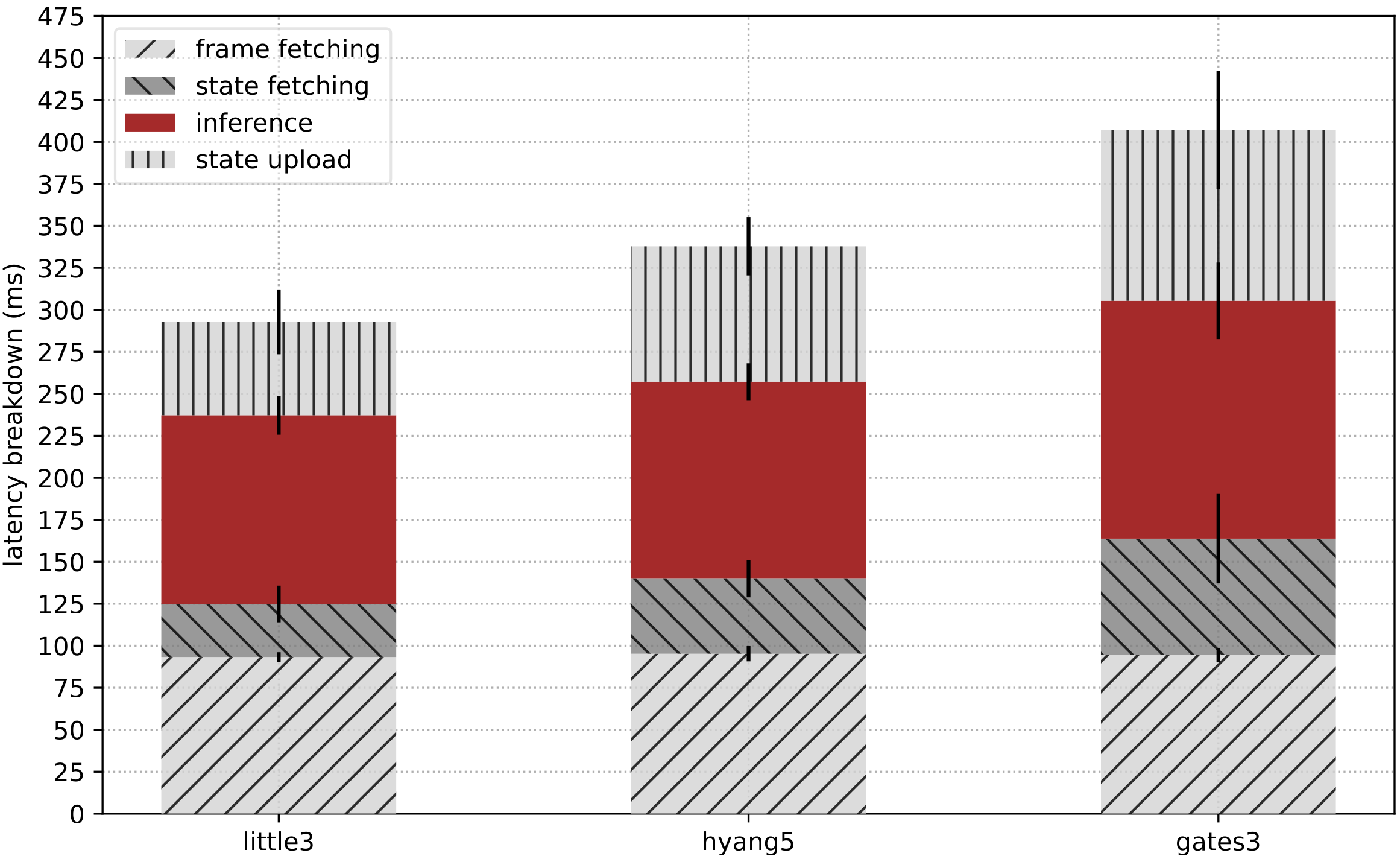}
  \caption{MOT latency breakdown.}
  \label{fig:mot_breakdown}
  \vspace{-5mm}
\end{figure}

\subsection{Increasing Workload}
\label{subsec:azure_1cam}

We now report results for experiments with just one client, a case that
establishes baseline latency for each video stream, processed individually in a
heavily provisioned,  dedicated infrastructure. In this set of experiments, each
step of the application had one corresponding endpoint, and we varied the number
of instances. The MOT endpoint had only one instance, while PRED and CD
endpoints had 1, 3, and 5.

Figure~\ref{fig:e2e_1cam} shows a box plot of the E2E latencies (in seconds) for
each video and number of PRED/CD instances. The number of instances for PRED and
CD endpoints is indicated on the horizontal axis. For \gates, the latency for 1
instance was significantly high and thus the bar was replaced by an arrow. The
plot shows that increasing the number of instances from 1 to 3 reduced the
median and 75th percentiles.  However, no significant benefit was observed when
increasing the number of instances to 5 for \little and \hyang. The reason for
this is that 3 instances were enough to significantly parallelize invocations to
PRED.  Increasing the number of instances even further resulted in more network
overhead of fetching actor positions that offset the gain of extra parallelism,
since the rate of cache misses increased. With 3 instances, 64ms was spent per
frame fetching data in the PRED step for \hyang, while 74ms was spent on average
with 5 instances.

We increased the workload to two simultaneous clients (\littlehyang),
using the same configuration as the experiments above for individual clients.
The average E2E latency, for 5 instances of PRED/CD, was about 67
seconds. The main reason for this significant increase in latency is that a
single instance for the MOT step is inadequate to handle the increased rate of
incoming frames (5 FPS in the aggregate from the two clients). It is necessary
to increase the number of instances of the MOT endpoint.

We conducted experiments with two simultaneous clients, employing 5 instances
for PRED and CD endpoints and a varying number of instances for the MOT endpoint
(3, 5, 7, and 9). For 3 MOT instances, the E2E latency was significantly
high. With 5 to 9 instances, the median latencies were all above 4 seconds, with
similar distributions. The benefit of increasing the number of instances was
limited, mainly due to the extra network overhead of fetching the MOT states.
The extra delay causes requests to pile up in queues, resulting in a
significantly higher E2E latency. Figure~\ref{fig:mot_breakdown} shows a
breakdown of the MOT step for each of the videos, when employing 3 instances on
the MOT endpoint. The breakdown shows how much time was spent on average (in ms)
on different MOT sub-steps: fetching the video frame, retrieving the state from
the previous frame, inference, and uploading the state of the current frame.

\begin{figure}[h!]
  \centering
  \includegraphics[width=0.85\linewidth]{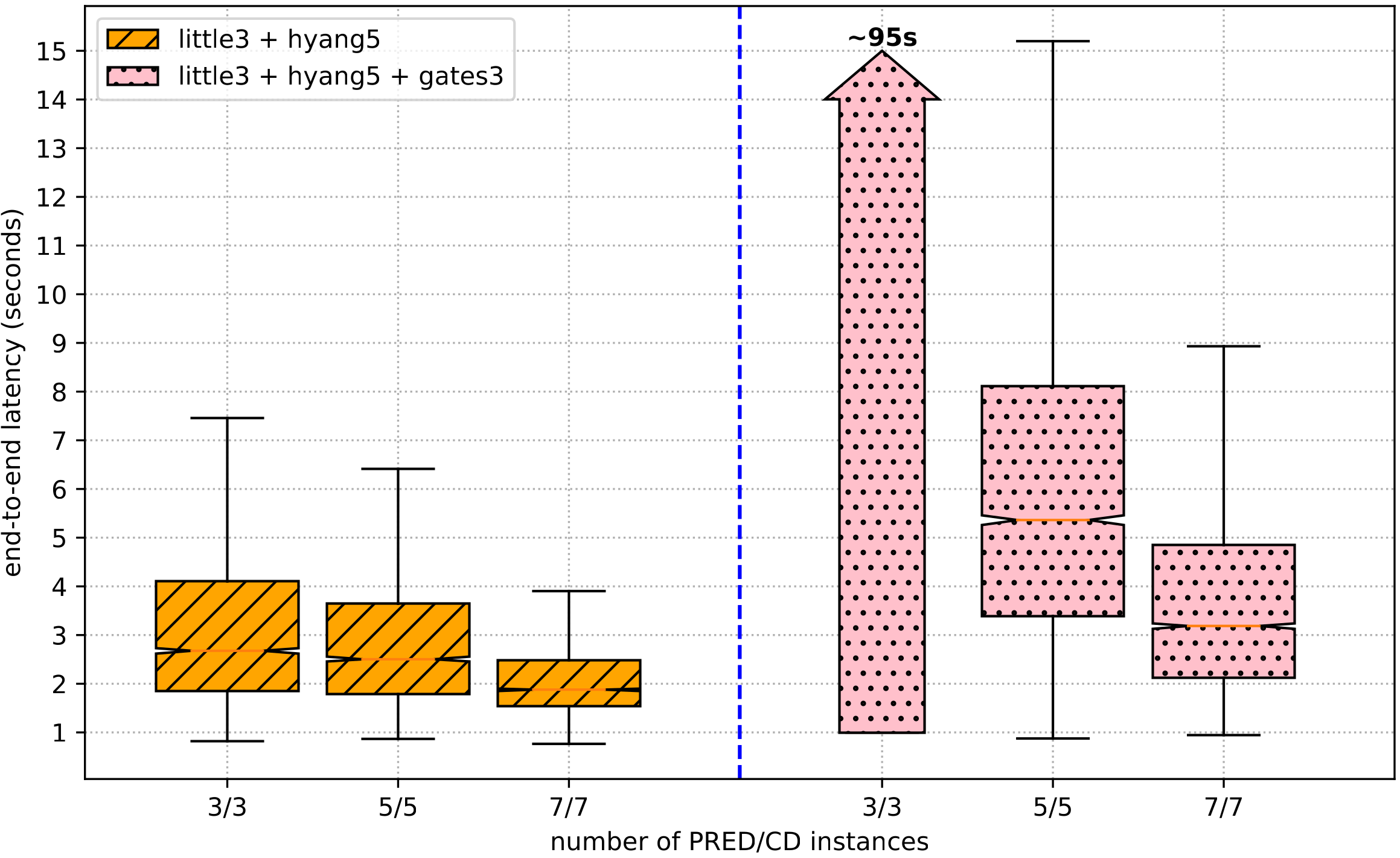}
  \caption{Grouped MOT with 2 and 3 clients.}
  \label{fig:grouped_mot_23cam}
  \vspace{-3mm}
\end{figure}

\subsection{Slashing Overheads From MOT}
\label{subsec:azure_mot}

Since we know that the MOT step requires state data regarding frames from the
same video, we deployed a separate MOT endpoint for each video, with one
instance each. A separate SA job was also deployed for each endpoint, receiving
requests from a separate EH. Then, each client sends requests to the
corresponding EH.  This ensures that frames from the same video will always be
received by the same instance, avoiding the overhead of fetching states since
they will be already in memory. We say that MOT step is now {\em grouped}.
Figure~\ref{fig:grouped_mot_23cam} shows the E2E latency of two and three
simultaneous clients, employing three MOT endpoints with one instance each, and
a varying number of PRED and CD instances (3, 5, and 7). It is possible to see
that latency became lower and more consistent as the number of instances for
PRED and CD endpoints increased.

\begin{figure}
  \centering
  \includegraphics[width=0.85\linewidth]{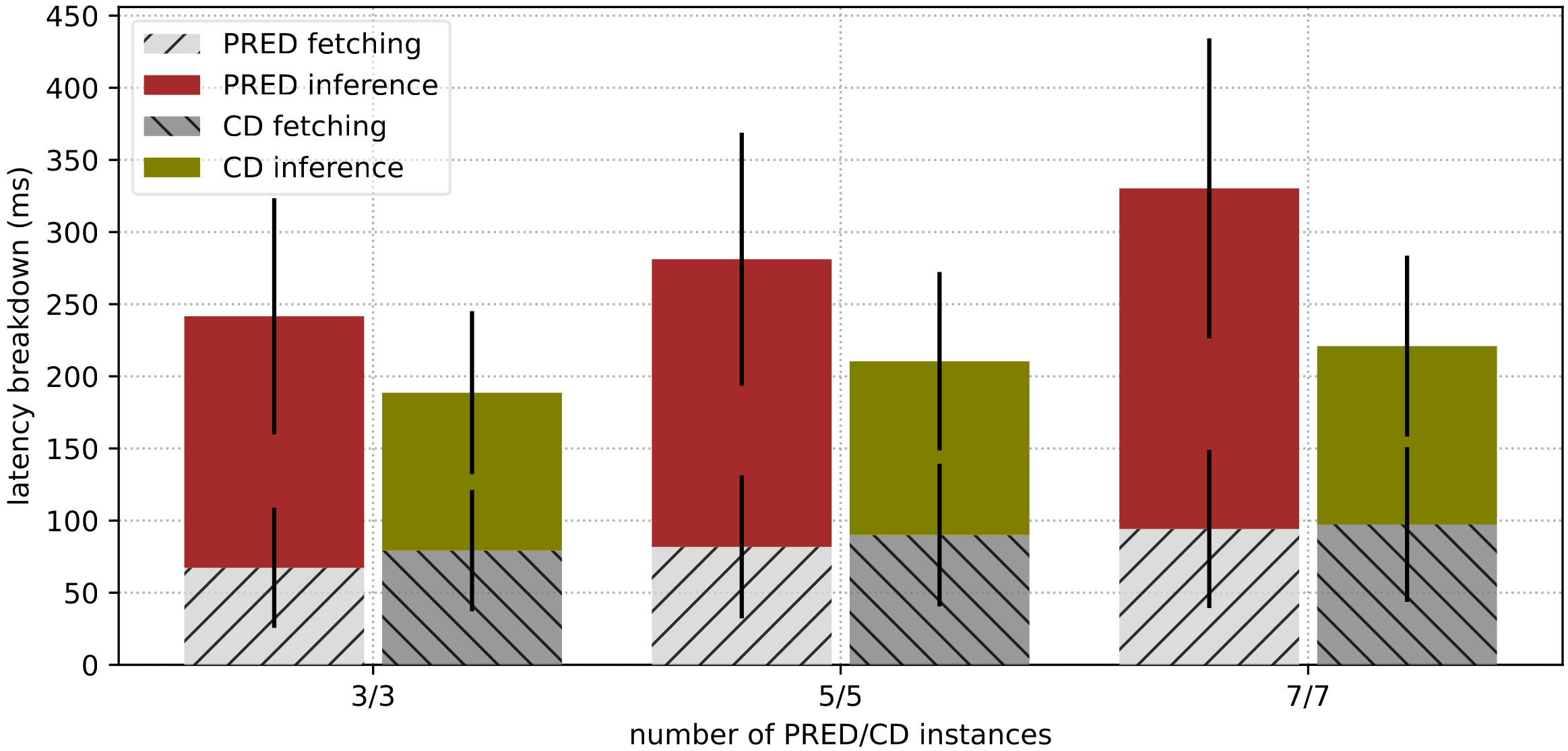}
  \caption{PRED and CD latency breakdown.}
  \label{fig:pred_cd_breakdown}
    \vspace{-5mm}
\end{figure}

However, we observed that there is still potential to further cut network
overheads. As we increased the number of PRED and CD instances, parallelism is
increased, but so is the time spent fetching actor positions and trajectory
predictions from Cosmos DB. Figure~\ref{fig:pred_cd_breakdown} shows the latency
breakdown for PRED and CD steps, with three simultaneous clients, three grouped
MOT endpoints, and a varying number of PRED and CD instances (3, 5, and 7). The
plot shows the average time spent per frame (in ms) fetching the necessary input
data and running the corresponding inference. The values shown are a sum of the
measured times for all invocations. With 7 PRED/CD instances, for example, the
average time spent fetching data per frame in PRED and CD steps was almost
200ms (roughly 100ms for each step).

\subsection{Slashing Overheads From PRED and CD}
\label{subsec:azure_predcd}

Here we employ the same approach as described above for MOT, leveraging
application-specific knowledge to group PRED and CD. We deploy multiple
endpoints with one instance each, instead of one endpoint with multiple
instances. However, it is not possible to rely any longer on the endpoint load
balancer to select which instance will receive each request. Thus, we manually
select in the SA jobs code which endpoint to forward each request, by writing
the output of the previous step to the corresponding EH. For PRED, this
selection was based on the actor identifier, modulo the number of PRED
endpoints. For CD, the selection was based on the frame number, modulo the
number of CD endpoints.

Listing~\ref{list:code_excerpt} shows a code excerpt (simplified for the sake of
presentation) from the SA job responsible for the MOT endpoint associated with
the video \little. In the code, {\em from-little3-client} is an EH where the
client responsible for the \little video sends requests. Function {\em
mot-little3-endpoint} invokes the AML endpoint containing a single instance that
performs the MOT inference. There are two queries (lines 9 and 12) that forwards
the position of each actor to a specific EH ({\em to-pred-0} or {\em to-pred-1})
depending on the actor identifier and the number of PRED endpoints. The EHs {\em
to-pred-0} and {\em to-pred-1} will deliver the actors positions to the
corresponding SA jobs that will invoke their associated PRED endpoints.

\definecolor{cadetblue}{rgb}{0.37, 0.62, 0.63}
\definecolor{darkolivegreen}{rgb}{0.33, 0.42, 0.18}

\begin{lstlisting}[
    caption={Code excerpt for the MOT SA job.},label=list:code_excerpt,captionpos=t,float,abovecaptionskip=-\medskipamount,
    linewidth=0.9\linewidth,
%    backgroundcolor=\color{white},
    language=SQL,
%    frame=single,
    numbers=right,
%    basicstyle=\scriptsize,
    basicstyle=\tiny,
    keywordstyle=\ttfamily\color{darkolivegreen},
    identifierstyle=\ttfamily\color{cadetblue}\bfseries,
    commentstyle=\color{brown},
    stringstyle=\ttfamily,
    showstringspaces=true]
WITH mot_output AS (
  SELECT [mot-little3-endpoint](input) AS res
  FROM [from-little3-client] AS input
), expanded_output AS (
  SELECT r.arrayvalue.* FROM mot_output
  CROSS APPLY GetArrayElements(res) AS r
)

SELECT * INTO [to-pred-0] FROM expanded_output
WHERE actor_id % num_pred_endpoints = 0

SELECT * INTO [to-pred-1] FROM expanded_output
WHERE actor_id % num_pred_endpoints = 1
\end{lstlisting}

\begin{figure}
  \centering
  \includegraphics[width=0.85\linewidth]{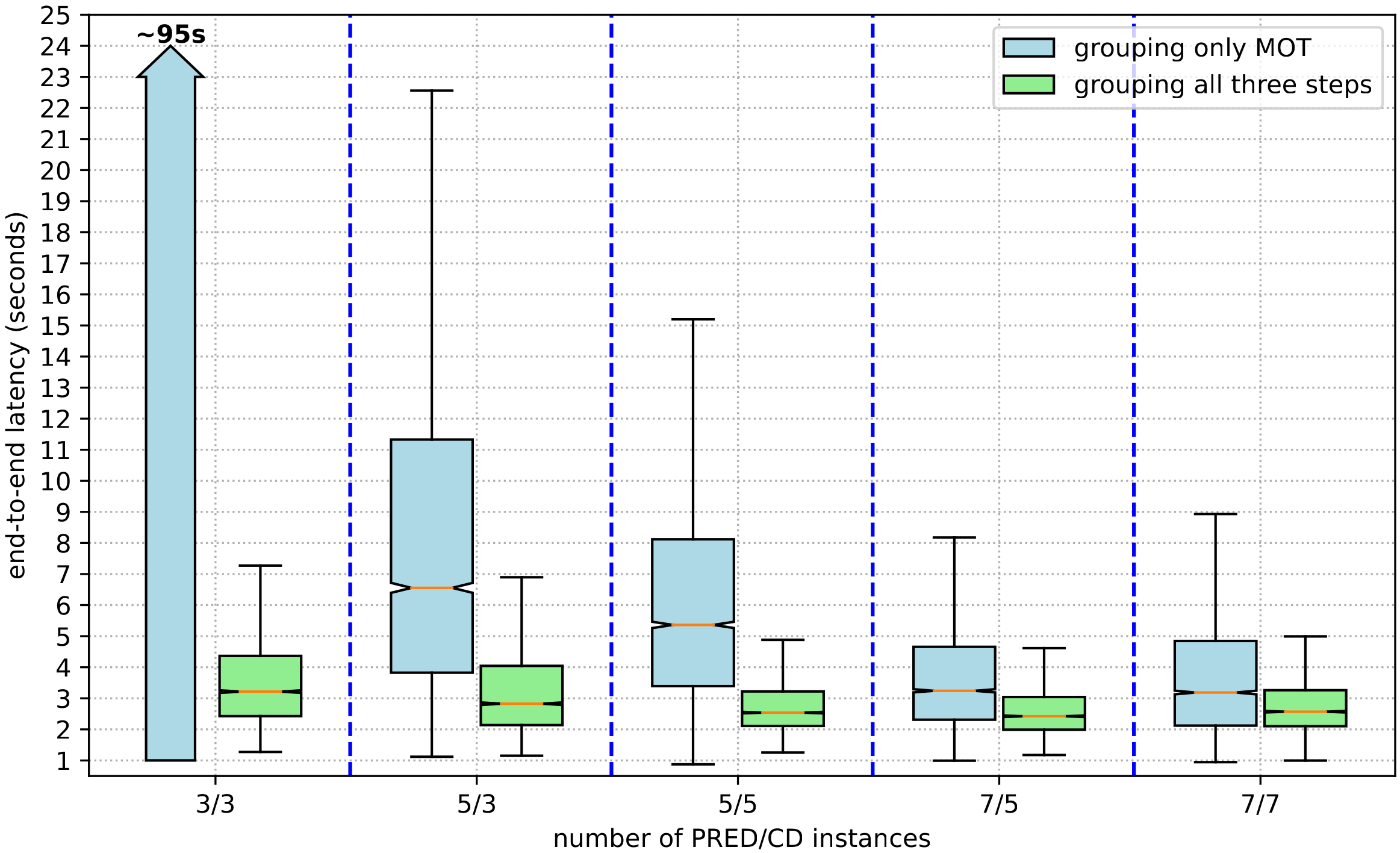}
  \caption{Grouping only MOT vs grouping all steps.}
  \label{fig:3cam_all_grouped}
    \vspace{-3mm}
\end{figure}

Figure~\ref{fig:3cam_all_grouped} shows a comparison between the pipeline with
only the MOT step grouped and the pipeline with all three steps grouped. We
varied the number of endpoints (with one instance each) for PRED and CD (3, 5,
and 7, in different combinations), while MOT always had three endpoints (one for
each video). As a workload, we configured three side-by-side client streams.
Latency for 3 PRED/CD instances and only MOT grouped was significantly high
(median 95 seconds) and was replaced by an arrow for the sake of presentation.
It is possible to observe a significant benefit when grouping all three steps.
The measured latency stabilizes with 5 PRED/CD instances, no benefit was
observed when further adding more instances. The gain in parallelism was
limited, however, we argue that more instances would be able to handle a heavier
workload without an increase in latency.

\subsection{Insights, Trade-offs, and Limitations}
\label{subsec:azure_tradeoffs}

Without our manual grouping, scaling in Azure is straightforward: it suffices
to increase the number of AML instances, SA streaming units
and/or EH partitions. AML endpoints will distribute requests across all the
available instances automatically according to their load. Furthermore, the
pattern lends itself to autoscaling.  In contrast, with manual grouping scaling entails adding
or removing endpoints, which requires that the application be reconfigured. In the
case of Azure, changing endpoints requires code changes to client and SA jobs
and the creation of new EHs, and the application itself would have to do load-balancing/auto-scaling.

The benefit of grouping is significant, echoing our observation 
in Section~\ref{sec:cascade}. For MOT, grouping
was essential to support more than one client.
Without grouping, the overhead of fetching states inflates runtimes to more than 400ms, meaning
that the next frame arrives before the current frame is processed (at 2.5 FPS). Requests will pile up at the first pipeline stage.
For a time-pressured scenario such as the RCP application, the
400ms threshold is of high importance: the AI model employed in the PRED step
predicts actors trajectories in the next 4.8 seconds. If it takes too long to
process each frame, due to queuing, it will be too late to act. Furthermore, since the application handles a real-time video stream, the
MOT state data for each frame will only be used once (by the subsequent frame), yet without grouping is 
moved twice (upload and download).


\section{Related Work}\label{sec:related}

Prior efforts~\cite{pretzel,willump,raven,mlnet,masq} argue for a {\em
white box} approach, in which optimizations are employed in compilation or
training time, based on specific features such as topology of neural networks.
While such approaches can reduce the computational cost of model serving, our
work employs a {\em black box} approach in which the application only needs to
identify data/compute correlations.  Although the authors in \cite{pretzel}
criticize {\em black box} approaches, they focus only on caching, buffering and
batching in non-pipelined applications.  The \afgrouping mechanism can support
all of these capabilities but also enables additional optimizations, such as
proactive data/computation collocation and prefetching.

Existing systems offer a number of mechanisms to express application-specific
knowledge and/or request grouping; these features can be found in Redis
\hashtag, Azure Cosmos DB, EHs and SA \partkey, RDD {\em location preferences},
Apache Spark Streaming {\em partitions}, and Apache Storm {\em stream grouping}.
However, as noted in Section~\ref{sec:affinity}, such mechanisms: (i) are highly
coupled with deployment; (ii) do not collocate data and computation in a unified
fashion; (iii) are not able to express all possible correlations between data
and computation; and/or (iv) may have a non-negligible computational cost as the
deployment scales out. 
We compare \afgrouping with these other approaches in more detail in Section~\ref{subsec:req}.

The Pheromone~\cite{pheromone} system offers a 
{\em data bucket} abstraction, in which the outputs of functions
are automatically grouped into buckets.  Developers can then arrange for
downstream functions to use data buckets as inputs. This enables Pheromone to
place functions close to where the data is stored. The data bucket abstraction
is similar to \afgrouping, however outputs in data buckets are always volatile
(once consumed, they are garbage collected).  The \afgrouping approach is
compatible with data persistence, and we would argue that this gives greater
flexibility.  
Furthermore, the authors do not consider more complex workflows in which
multiple functions may require access to the same output in different stages, or
in which a function requires data from multiple buckets. 

Several works improve data locality by scheduling tasks where input data (or
most of it) is
located~\cite{rtml,ray,cloudburst,chukonu,apollo,inferline,videostorm}, or by
scheduling tasks that interact with each other in the same location (node, VM,
or container)~\cite{orion,faastlane}.  These solutions follow a reactive
approach: data is first placed without taking into account their correlations,
and then tasks are placed where ``most of input data'' is located.  Due to the
data access patterns of latency-sensitive AI pipelines, a proactive approach,
such as enabled by \afgrouping, may be necessary. We claim that our proposed
mechanism is complementary to scheduler-based solutions.

Finally, data collocation is widely explored in literature and thus is not a
novel concept per se~\cite{bigdata_survey}. However, we argue that our novelty
lies in providing developers with an easy-to-use mechanism that requires no
knowledge of deployment/environment details, while still achieving effective
collocation.  The experiments in Appendix~\ref{sec:azure} are evidence of the
limitation of modern platforms and the potential to improve latency by applying
collocation principles. Nevertheless, achieving such collocation is challenging for
developers and pose trade-offs, hence the need for a better mechanism.

\section{Beyond RCP and Cascade}\label{sec:beyond}

In this work we employed the RCP application as a running example, and implemented \afgrouping only
on the Cascade platform.  A natural question that arises is whether \afgrouping can be generalized
to other applications and/or AI hosting platforms. To address such question, we discuss next how the
\afgrouping mechanism could be implemented in Azure, as well as other emerging applications where
\afgrouping can have a great impact.

\subsection{\Afgrouping in Azure}

In Section~\ref{sec:cascade} we described an implementation on a platform based on a K/V store. Few
modifications would be required to implement our proposals in Azure.  For example, if Azure's
storage solutions and AML operated in a consistent fashion, data stored in Cosmos DB or Blob storage
with a certain \afkey could be prefetched by physical servers running the AML instance handling
requests with that \afkey.  AML endpoints with multiple instances should ideally route requests with
the same \afkey to the same instance. If the load balancer will add instances to reduce load on a
hot-spot, it could prefetch correlated data to pre-warm those instances.  Azure has many caching
components; these could be extended to load or evict object sets, applying the identical policy
objects sharing a single \afkey.

\subsection{Emerging applications}

In reinforcement learning (RL)~\cite{rlhf}, an ML system might run for hours or days repeatedly accessing the
same data objects, but not always using the same nodes for running tasks. A developer of such system
knows which data objects are accessed by each task, and therefore can map \afkeys to tasks and data
objects. Such mapping allows a node to fetch all needed objects for a task invocation (they all
share the same \afkey) at once and in parallel, and cache them as a set (or evict them as a set).
Such objects may be a set of cached K/V pairs, corresponding to the activation state of the training
NN, needed by a Low Rank Adapter (LoRA). Fetching one by one incurs significantly higher delays,
because it doesn't leverage the parallelism of the network, and caching systems lack semantic
knowledge of the application, which is clear for the developer and can be easily encoded through
\afgrouping.

The same rationale above also applies in the training of modern transformer-based models (e.g.
LLMs). In such context, it is common practice to cache collections of Query ($Q$), Key ($K$), Value
($V$) tuples, which are a snapshot of the DNN state of some layer $L$ at step $T$, triggered by
input query Q.  Any given ($Q, K, V$) tuple is updated by only one neuron in layer $L$,  but the
iterated (``auto-regressive'') training algorithm requires that at each step every neuron in layer
$L$ read the full set of layer $L$ tuples from the prior step, adapt its own weights (compute a
gradient), apply the gradient to the layers around it (forward and back propagation), and then
update its (Q, K, V) tuple. 
By hashing $Q$, $L$ and $T$ we can identify an affinity group:  the entire group of tuples for this
query at this time step for layer $L$ can be cached (or evicted) jointly.  Updates will stream from
individual neurons to the tuples they own, and as each step finishes, the neurons active in the next
step will simply read the cached prior values.

In contrast, suppose we viewed this affinity grouping as a key in a standard K/V store like Reddis.
Clearly, we could form a single object collecting the set of ($Q, K, V$) tuples for layer $L$ at
step $T$.  Once serialized to a byte vector, we would have a K/V object that can be stored in a
sharded key-value storage service.  But because this one big object has contributions from all
neurons in $L$, we would either need to run AllReduce simply to compute it, or pick a K/V store that
supports coherent in-place updates and have each neuron read the working value, update it to add its
own ($Q, K, V$) tuple, and write the result back. After one update per neuron, the version of the
object would be complete for step $T$, and step $T+1$ could run. The issue here is that all the
neurons in layer $L$ contribute to this object, and because they run concurrently, all contend to
read, modify, and rewrite the object when creating it. That will take time: one layer can have many
neurons, so these two alternative approaches (AllReduce or one big object holding the collection)
would be very slow in comparison to an affinity-based solution, where each node only reads tuples
from its own cache.

\section{Conclusion}\label{sec:conc}

We proposed an \afgrouping mechanism that enables developers to express application-specific knowledge
of data/computation correlations, and then for platforms to use this information to reduce
latency and improve efficiency.  Our results show
that the proposed mechanism is able to maintain significantly lower latency as the application
workload increases and the infrastructure scales out. Results also show great potential for integrating the proposed mechanism with 
other approaches such as prefetching and scheduling.



\bibliographystyle{ACM-Reference-Format}
\bibliography{refs}


\begin{thebibliography}{41}


\ifx \showCODEN    \undefined \def \showCODEN     #1{\unskip}     \fi
\ifx \showISBNx    \undefined \def \showISBNx     #1{\unskip}     \fi
\ifx \showISBNxiii \undefined \def \showISBNxiii  #1{\unskip}     \fi
\ifx \showISSN     \undefined \def \showISSN      #1{\unskip}     \fi
\ifx \showLCCN     \undefined \def \showLCCN      #1{\unskip}     \fi
\ifx \shownote     \undefined \def \shownote      #1{#1}          \fi
\ifx \showarticletitle \undefined \def \showarticletitle #1{#1}   \fi
\ifx \showURL      \undefined \def \showURL       {\relax}        \fi
\providecommand\bibfield[2]{#2}
\providecommand\bibinfo[2]{#2}
\providecommand\natexlab[1]{#1}
\providecommand\showeprint[2][]{arXiv:#2}

\bibitem[Ahmed et~al\mbox{.}(2019)]%
        {mlnet}
\bibfield{author}{\bibinfo{person}{Zeeshan Ahmed}, \bibinfo{person}{Saeed
  Amizadeh}, \bibinfo{person}{Mikhail Bilenko}, \bibinfo{person}{Rogan Carr},
  \bibinfo{person}{Wei-Sheng Chin}, \bibinfo{person}{Yael Dekel},
  \bibinfo{person}{Xavier Dupre}, \bibinfo{person}{Vadim Eksarevskiy},
  \bibinfo{person}{Senja Filipi}, \bibinfo{person}{Tom Finley},
  \bibinfo{person}{Abhishek Goswami}, \bibinfo{person}{Monte Hoover},
  \bibinfo{person}{Scott Inglis}, \bibinfo{person}{Matteo Interlandi},
  \bibinfo{person}{Najeeb Kazmi}, \bibinfo{person}{Gleb Krivosheev},
  \bibinfo{person}{Pete Luferenko}, \bibinfo{person}{Ivan Matantsev},
  \bibinfo{person}{Sergiy Matusevych}, \bibinfo{person}{Shahab Moradi},
  \bibinfo{person}{Gani Nazirov}, \bibinfo{person}{Justin Ormont},
  \bibinfo{person}{Gal Oshri}, \bibinfo{person}{Artidoro Pagnoni},
  \bibinfo{person}{Jignesh Parmar}, \bibinfo{person}{Prabhat Roy},
  \bibinfo{person}{Mohammad~Zeeshan Siddiqui}, \bibinfo{person}{Markus Weimer},
  \bibinfo{person}{Shauheen Zahirazami}, {and} \bibinfo{person}{Yiwen Zhu}.}
  \bibinfo{year}{2019}\natexlab{}.
\newblock \showarticletitle{Machine Learning at Microsoft with ML.NET}. In
  \bibinfo{booktitle}{\emph{Proceedings of the 25th ACM SIGKDD International
  Conference on Knowledge Discovery \& Data Mining}} (Anchorage, AK, USA)
  \emph{(\bibinfo{series}{KDD '19})}. \bibinfo{publisher}{Association for
  Computing Machinery}, \bibinfo{address}{New York, NY, USA},
  \bibinfo{pages}{2448–2458}.
\newblock
\showISBNx{9781450362016}
\href{https://doi.org/10.1145/3292500.3330667}{doi:\nolinkurl{10.1145/3292500.3330667}}


\bibitem[Boutin et~al\mbox{.}(2014)]%
        {apollo}
\bibfield{author}{\bibinfo{person}{Eric Boutin}, \bibinfo{person}{Jaliya
  Ekanayake}, \bibinfo{person}{Wei Lin}, \bibinfo{person}{Bing Shi},
  \bibinfo{person}{Jingren Zhou}, \bibinfo{person}{Zhengping Qian},
  \bibinfo{person}{Ming Wu}, {and} \bibinfo{person}{Lidong Zhou}.}
  \bibinfo{year}{2014}\natexlab{}.
\newblock \showarticletitle{Apollo: Scalable and Coordinated Scheduling for
  {Cloud-Scale} Computing}. In \bibinfo{booktitle}{\emph{11th USENIX Symposium
  on Operating Systems Design and Implementation (OSDI 14)}}.
  \bibinfo{publisher}{USENIX Association}, \bibinfo{address}{Broomfield, CO},
  \bibinfo{pages}{285--300}.
\newblock
\showISBNx{978-1-931971-16-4}
\urldef\tempurl%
\url{https://www.usenix.org/conference/osdi14/technical-sessions/presentation/boutin}
\showURL{%
\tempurl}


\bibitem[Broström(2022)]%
        {yolov5-strongsort-osnet-2022}
\bibfield{author}{\bibinfo{person}{Mikel Broström}.}
  \bibinfo{year}{2022}\natexlab{}.
\newblock \bibinfo{booktitle}{\emph{Real-time multi-camera multi-object tracker
  using {YOLOv5} and {StrongSORT} with {OSNet}}}.
\newblock
\urldef\tempurl%
\url{https://github.com/mikel-brostrom/Yolov5_StrongSORT_OSNet}
\showURL{%
\tempurl}


\bibitem[Calder et~al\mbox{.}(2011)]%
        {calder2011windows}
\bibfield{author}{\bibinfo{person}{Brad Calder}, \bibinfo{person}{Ju Wang},
  \bibinfo{person}{Aaron Ogus}, \bibinfo{person}{Niranjan Nilakantan},
  \bibinfo{person}{Arild Skjolsvold}, \bibinfo{person}{Sam McKelvie},
  \bibinfo{person}{Yikang Xu}, \bibinfo{person}{Shashwat Srivastav},
  \bibinfo{person}{Jiesheng Wu}, \bibinfo{person}{Huseyin Simitci},
  {et~al\mbox{.}}} \bibinfo{year}{2011}\natexlab{}.
\newblock \showarticletitle{Windows azure storage: a highly available cloud
  storage service with strong consistency}. In
  \bibinfo{booktitle}{\emph{Proceedings of the Twenty-Third ACM Symposium on
  Operating Systems Principles}}. \bibinfo{pages}{143--157}.
\newblock


\bibitem[Chen et~al\mbox{.}(2021)]%
        {olda}
\bibfield{author}{\bibinfo{person}{Cheng Chen}, \bibinfo{person}{Jun Yang},
  \bibinfo{person}{Mian Lu}, \bibinfo{person}{Taize Wang},
  \bibinfo{person}{Zhao Zheng}, \bibinfo{person}{Yuqiang Chen},
  \bibinfo{person}{Wenyuan Dai}, \bibinfo{person}{Bingsheng He},
  \bibinfo{person}{Weng-Fai Wong}, \bibinfo{person}{Guoan Wu},
  \bibinfo{person}{Yuping Zhao}, {and} \bibinfo{person}{Andy Rudoff}.}
  \bibinfo{year}{2021}\natexlab{}.
\newblock \showarticletitle{Optimizing In-Memory Database Engine for AI-Powered
  on-Line Decision Augmentation Using Persistent Memory}.
\newblock \bibinfo{journal}{\emph{Proc. VLDB Endow.}} \bibinfo{volume}{14},
  \bibinfo{number}{5} (\bibinfo{date}{jan} \bibinfo{year}{2021}),
  \bibinfo{pages}{799–812}.
\newblock
\showISSN{2150-8097}
\href{https://doi.org/10.14778/3446095.3446102}{doi:\nolinkurl{10.14778/3446095.3446102}}


\bibitem[Crankshaw et~al\mbox{.}(2020)]%
        {inferline}
\bibfield{author}{\bibinfo{person}{Daniel Crankshaw}, \bibinfo{person}{Gur-Eyal
  Sela}, \bibinfo{person}{Xiangxi Mo}, \bibinfo{person}{Corey Zumar},
  \bibinfo{person}{Ion Stoica}, \bibinfo{person}{Joseph Gonzalez}, {and}
  \bibinfo{person}{Alexey Tumanov}.} \bibinfo{year}{2020}\natexlab{}.
\newblock \showarticletitle{InferLine: Latency-Aware Provisioning and Scaling
  for Prediction Serving Pipelines}. In \bibinfo{booktitle}{\emph{Proceedings
  of the 11th ACM Symposium on Cloud Computing}} (Virtual Event, USA)
  \emph{(\bibinfo{series}{SoCC '20})}. \bibinfo{publisher}{Association for
  Computing Machinery}, \bibinfo{address}{New York, NY, USA},
  \bibinfo{pages}{477–491}.
\newblock
\showISBNx{9781450381376}
\href{https://doi.org/10.1145/3419111.3421285}{doi:\nolinkurl{10.1145/3419111.3421285}}


\bibitem[Crankshaw et~al\mbox{.}(2017)]%
        {clipper}
\bibfield{author}{\bibinfo{person}{Daniel Crankshaw}, \bibinfo{person}{Xin
  Wang}, \bibinfo{person}{Guilio Zhou}, \bibinfo{person}{Michael~J. Franklin},
  \bibinfo{person}{Joseph~E. Gonzalez}, {and} \bibinfo{person}{Ion Stoica}.}
  \bibinfo{year}{2017}\natexlab{}.
\newblock \showarticletitle{Clipper: A {Low-Latency} Online Prediction Serving
  System}. In \bibinfo{booktitle}{\emph{14th USENIX Symposium on Networked
  Systems Design and Implementation (NSDI 17)}}. \bibinfo{publisher}{USENIX
  Association}, \bibinfo{address}{Boston, MA}, \bibinfo{pages}{613--627}.
\newblock
\showISBNx{978-1-931971-37-9}
\urldef\tempurl%
\url{https://www.usenix.org/conference/nsdi17/technical-sessions/presentation/crankshaw}
\showURL{%
\tempurl}


\bibitem[Deng et~al\mbox{.}(2020)]%
        {edgeint}
\bibfield{author}{\bibinfo{person}{Shuiguang Deng}, \bibinfo{person}{Hailiang
  Zhao}, \bibinfo{person}{Weijia Fang}, \bibinfo{person}{Jianwei Yin},
  \bibinfo{person}{Schahram Dustdar}, {and} \bibinfo{person}{Albert~Y.
  Zomaya}.} \bibinfo{year}{2020}\natexlab{}.
\newblock \showarticletitle{Edge Intelligence: The Confluence of Edge Computing
  and Artificial Intelligence}.
\newblock \bibinfo{journal}{\emph{IEEE Internet of Things Journal}}
  \bibinfo{volume}{7}, \bibinfo{number}{8} (\bibinfo{year}{2020}),
  \bibinfo{pages}{7457--7469}.
\newblock
\href{https://doi.org/10.1109/JIOT.2020.2984887}{doi:\nolinkurl{10.1109/JIOT.2020.2984887}}


\bibitem[Du et~al\mbox{.}(2022)]%
        {strongsort}
\bibfield{author}{\bibinfo{person}{Yunhao Du}, \bibinfo{person}{Yang Song},
  \bibinfo{person}{Bo Yang}, {and} \bibinfo{person}{Yanyun Zhao}.}
  \bibinfo{year}{2022}\natexlab{}.
\newblock \showarticletitle{Strongsort: Make Deepsort Great Again}.
\newblock \bibinfo{journal}{\emph{arXiv preprint arXiv:2202.13514}}
  (\bibinfo{year}{2022}).
\newblock


\bibitem[Eidson et~al\mbox{.}(2002)]%
        {ptp}
\bibfield{author}{\bibinfo{person}{John~C Eidson}, \bibinfo{person}{Mike
  Fischer}, {and} \bibinfo{person}{Joe White}.}
  \bibinfo{year}{2002}\natexlab{}.
\newblock \showarticletitle{{IEEE-1588™} Standard for a Precision Clock
  Synchronization Protocol for Networked Measurement and Control Systems}. In
  \bibinfo{booktitle}{\emph{Proceedings of the 34th Annual Precise Time and
  Time Interval Systems and Applications Meeting}}. \bibinfo{pages}{243--254}.
\newblock


\bibitem[Girase(2022)]%
        {ynetCode}
\bibfield{author}{\bibinfo{person}{Harshayu Girase}.}
  \bibinfo{year}{2022}\natexlab{}.
\newblock \bibinfo{booktitle}{\emph{Human Path Prediction}}.
\newblock
\urldef\tempurl%
\url{https://github.com/HarshayuGirase/Human-Path-Prediction}
\showURL{%
\tempurl}


\bibitem[Isah et~al\mbox{.}(2019)]%
        {stream}
\bibfield{author}{\bibinfo{person}{Haruna Isah}, \bibinfo{person}{Tariq
  Abughofa}, \bibinfo{person}{Sazia Mahfuz}, \bibinfo{person}{Dharmitha
  Ajerla}, \bibinfo{person}{Farhana Zulkernine}, {and} \bibinfo{person}{Shahzad
  Khan}.} \bibinfo{year}{2019}\natexlab{}.
\newblock \showarticletitle{{A Survey of Distributed Data Stream Processing
  Frameworks}}.
\newblock \bibinfo{journal}{\emph{IEEE Access}}  \bibinfo{volume}{7}
  (\bibinfo{year}{2019}), \bibinfo{pages}{154300--154316}.
\newblock
\href{https://doi.org/10.1109/ACCESS.2019.2946884}{doi:\nolinkurl{10.1109/ACCESS.2019.2946884}}


\bibitem[Kettner and Geisler(2022)]%
        {kettner2022iot}
\bibfield{author}{\bibinfo{person}{Benjamin Kettner} {and}
  \bibinfo{person}{Frank Geisler}.} \bibinfo{year}{2022}\natexlab{}.
\newblock \showarticletitle{IoT Hub, Event Hub, and Streaming Data}.
\newblock In \bibinfo{booktitle}{\emph{Pro Serverless Data Handling with
  Microsoft Azure: Architecting ETL and Data-Driven Applications in the
  Cloud}}. \bibinfo{publisher}{Springer}, \bibinfo{pages}{153--168}.
\newblock


\bibitem[Kotni et~al\mbox{.}(2021)]%
        {faastlane}
\bibfield{author}{\bibinfo{person}{Swaroop Kotni}, \bibinfo{person}{Ajay
  Nayak}, \bibinfo{person}{Vinod Ganapathy}, {and} \bibinfo{person}{Arkaprava
  Basu}.} \bibinfo{year}{2021}\natexlab{}.
\newblock \showarticletitle{Faastlane: Accelerating {Function-as-a-Service}
  Workflows}. In \bibinfo{booktitle}{\emph{2021 USENIX Annual Technical
  Conference (USENIX ATC 21)}}. \bibinfo{publisher}{USENIX Association},
  \bibinfo{pages}{805--820}.
\newblock
\showISBNx{978-1-939133-23-6}
\urldef\tempurl%
\url{https://www.usenix.org/conference/atc21/presentation/kotni}
\showURL{%
\tempurl}


\bibitem[Kraft et~al\mbox{.}(2020)]%
        {willump}
\bibfield{author}{\bibinfo{person}{Peter Kraft}, \bibinfo{person}{Daniel Kang},
  \bibinfo{person}{Deepak Narayanan}, \bibinfo{person}{Shoumik Palkar},
  \bibinfo{person}{Peter Bailis}, {and} \bibinfo{person}{Matei Zaharia}.}
  \bibinfo{year}{2020}\natexlab{}.
\newblock \showarticletitle{Willump: A Statistically-Aware End-to-end Optimizer
  for Machine Learning Inference}. In \bibinfo{booktitle}{\emph{Proceedings of
  Machine Learning and Systems}},
  \bibfield{editor}{\bibinfo{person}{I.~Dhillon},
  \bibinfo{person}{D.~Papailiopoulos}, {and} \bibinfo{person}{V.~Sze}} (Eds.),
  Vol.~\bibinfo{volume}{2}. \bibinfo{pages}{147--159}.
\newblock
\urldef\tempurl%
\url{https://proceedings.mlsys.org/paper_files/paper/2020/file/d9e5bd751997cffa6bc2d0e31ebdc048-Paper.pdf}
\showURL{%
\tempurl}


\bibitem[Lee et~al\mbox{.}(2018)]%
        {pretzel}
\bibfield{author}{\bibinfo{person}{Yunseong Lee}, \bibinfo{person}{Alberto
  Scolari}, \bibinfo{person}{Byung-Gon Chun}, \bibinfo{person}{Marco~Domenico
  Santambrogio}, \bibinfo{person}{Markus Weimer}, {and} \bibinfo{person}{Matteo
  Interlandi}.} \bibinfo{year}{2018}\natexlab{}.
\newblock \showarticletitle{{PRETZEL}: Opening the Black Box of Machine
  Learning Prediction Serving Systems}. In \bibinfo{booktitle}{\emph{13th
  USENIX Symposium on Operating Systems Design and Implementation (OSDI 18)}}.
  \bibinfo{publisher}{USENIX Association}, \bibinfo{address}{Carlsbad, CA},
  \bibinfo{pages}{611--626}.
\newblock
\showISBNx{978-1-939133-08-3}
\urldef\tempurl%
\url{https://www.usenix.org/conference/osdi18/presentation/lee}
\showURL{%
\tempurl}


\bibitem[Mahgoub et~al\mbox{.}(2022)]%
        {orion}
\bibfield{author}{\bibinfo{person}{Ashraf Mahgoub},
  \bibinfo{person}{Edgardo~Barsallo Yi}, \bibinfo{person}{Karthick Shankar},
  \bibinfo{person}{Sameh Elnikety}, \bibinfo{person}{Somali Chaterji}, {and}
  \bibinfo{person}{Saurabh Bagchi}.} \bibinfo{year}{2022}\natexlab{}.
\newblock \showarticletitle{{ORION} and the Three Rights: Sizing, Bundling, and
  Prewarming for Serverless {DAGs}}. In \bibinfo{booktitle}{\emph{16th USENIX
  Symposium on Operating Systems Design and Implementation (OSDI 22)}}.
  \bibinfo{publisher}{USENIX Association}, \bibinfo{address}{Carlsbad, CA},
  \bibinfo{pages}{303--320}.
\newblock
\showISBNx{978-1-939133-28-1}
\urldef\tempurl%
\url{https://www.usenix.org/conference/osdi22/presentation/mahgoub}
\showURL{%
\tempurl}


\bibitem[Makrynioti and Vassalos(2021)]%
        {declarative}
\bibfield{author}{\bibinfo{person}{Nantia Makrynioti} {and}
  \bibinfo{person}{Vasilis Vassalos}.} \bibinfo{year}{2021}\natexlab{}.
\newblock \showarticletitle{Declarative Data Analytics: A Survey}.
\newblock \bibinfo{journal}{\emph{IEEE Transactions on Knowledge and Data
  Engineering}} \bibinfo{volume}{33}, \bibinfo{number}{6}
  (\bibinfo{year}{2021}), \bibinfo{pages}{2392--2411}.
\newblock
\href{https://doi.org/10.1109/TKDE.2019.2958084}{doi:\nolinkurl{10.1109/TKDE.2019.2958084}}


\bibitem[Mangalam et~al\mbox{.}(2021)]%
        {ynet}
\bibfield{author}{\bibinfo{person}{Karttikeya Mangalam}, \bibinfo{person}{Yang
  An}, \bibinfo{person}{Harshayu Girase}, {and} \bibinfo{person}{Jitendra
  Malik}.} \bibinfo{year}{2021}\natexlab{}.
\newblock \showarticletitle{{From Goals, Waypoints \& Paths to Long Term Human
  Trajectory Forecasting}}. In \bibinfo{booktitle}{\emph{Proceedings of the
  IEEE/CVF International Conference on Computer Vision (ICCV)}}.
  \bibinfo{pages}{15233--15242}.
\newblock


\bibitem[Mazumdar et~al\mbox{.}(2019)]%
        {bigdata_survey}
\bibfield{author}{\bibinfo{person}{Somnath Mazumdar}, \bibinfo{person}{Daniel
  Seybold}, \bibinfo{person}{Kyriakos Kritikos}, {and} \bibinfo{person}{Yiannis
  Verginadis}.} \bibinfo{year}{2019}\natexlab{}.
\newblock \showarticletitle{A survey on data storage and placement
  methodologies for cloud-big data ecosystem}.
\newblock \bibinfo{journal}{\emph{Journal of Big Data}} \bibinfo{volume}{6},
  \bibinfo{number}{1} (\bibinfo{year}{2019}), \bibinfo{pages}{1--37}.
\newblock


\bibitem[Microsoft(2022a)]%
        {asaandml}
\bibfield{author}{\bibinfo{person}{Microsoft}.}
  \bibinfo{year}{2022}\natexlab{a}.
\newblock \bibinfo{booktitle}{\emph{Integrate Azure Stream Analytics with Azure
  Machine Learning}}.
\newblock
\urldef\tempurl%
\url{https://learn.microsoft.com/en-us/azure/stream-analytics/machine-learning-udf}
\showURL{%
\tempurl}


\bibitem[Microsoft(2022b)]%
        {azureParallel}
\bibfield{author}{\bibinfo{person}{Microsoft}.}
  \bibinfo{year}{2022}\natexlab{b}.
\newblock \bibinfo{booktitle}{\emph{Leverage query parallelization in Azure
  Stream Analytics}}.
\newblock
\urldef\tempurl%
\url{https://learn.microsoft.com/en-us/azure/stream-analytics/stream-analytics-parallelization}
\showURL{%
\tempurl}


\bibitem[Moritz et~al\mbox{.}(2018)]%
        {ray}
\bibfield{author}{\bibinfo{person}{Philipp Moritz}, \bibinfo{person}{Robert
  Nishihara}, \bibinfo{person}{Stephanie Wang}, \bibinfo{person}{Alexey
  Tumanov}, \bibinfo{person}{Richard Liaw}, \bibinfo{person}{Eric Liang},
  \bibinfo{person}{Melih Elibol}, \bibinfo{person}{Zongheng Yang},
  \bibinfo{person}{William Paul}, \bibinfo{person}{Michael~I. Jordan}, {and}
  \bibinfo{person}{Ion Stoica}.} \bibinfo{year}{2018}\natexlab{}.
\newblock \showarticletitle{Ray: A Distributed Framework for Emerging {AI}
  Applications}. In \bibinfo{booktitle}{\emph{13th USENIX Symposium on
  Operating Systems Design and Implementation (OSDI 18)}}.
  \bibinfo{publisher}{USENIX Association}, \bibinfo{address}{Carlsbad, CA},
  \bibinfo{pages}{561--577}.
\newblock
\showISBNx{978-1-939133-08-3}
\urldef\tempurl%
\url{https://www.usenix.org/conference/osdi18/presentation/moritz}
\showURL{%
\tempurl}


\bibitem[Nishihara et~al\mbox{.}(2017)]%
        {rtml}
\bibfield{author}{\bibinfo{person}{Robert Nishihara}, \bibinfo{person}{Philipp
  Moritz}, \bibinfo{person}{Stephanie Wang}, \bibinfo{person}{Alexey Tumanov},
  \bibinfo{person}{William Paul}, \bibinfo{person}{Johann Schleier-Smith},
  \bibinfo{person}{Richard Liaw}, \bibinfo{person}{Mehrdad Niknami},
  \bibinfo{person}{Michael~I. Jordan}, {and} \bibinfo{person}{Ion Stoica}.}
  \bibinfo{year}{2017}\natexlab{}.
\newblock \showarticletitle{Real-Time Machine Learning: The Missing Pieces}. In
  \bibinfo{booktitle}{\emph{Proceedings of the 16th Workshop on Hot Topics in
  Operating Systems}} (Whistler, BC, Canada) \emph{(\bibinfo{series}{HotOS
  '17})}. \bibinfo{publisher}{Association for Computing Machinery},
  \bibinfo{address}{New York, NY, USA}, \bibinfo{pages}{106–110}.
\newblock
\showISBNx{9781450350686}
\href{https://doi.org/10.1145/3102980.3102998}{doi:\nolinkurl{10.1145/3102980.3102998}}


\bibitem[Paganelli et~al\mbox{.}(2023)]%
        {masq}
\bibfield{author}{\bibinfo{person}{Matteo Paganelli}, \bibinfo{person}{Paolo
  Sottovia}, \bibinfo{person}{Kwanghyun Park}, \bibinfo{person}{Matteo
  Interlandi}, {and} \bibinfo{person}{Francesco Guerra}.}
  \bibinfo{year}{2023}\natexlab{}.
\newblock \showarticletitle{Pushing ML Predictions Into DBMSs}.
\newblock \bibinfo{journal}{\emph{IEEE Transactions on Knowledge and Data
  Engineering}} \bibinfo{volume}{35}, \bibinfo{number}{10}
  (\bibinfo{year}{2023}), \bibinfo{pages}{10295--10308}.
\newblock
\href{https://doi.org/10.1109/TKDE.2023.3269592}{doi:\nolinkurl{10.1109/TKDE.2023.3269592}}


\bibitem[Park et~al\mbox{.}(2022)]%
        {raven}
\bibfield{author}{\bibinfo{person}{Kwanghyun Park}, \bibinfo{person}{Karla
  Saur}, \bibinfo{person}{Dalitso Banda}, \bibinfo{person}{Rathijit Sen},
  \bibinfo{person}{Matteo Interlandi}, {and} \bibinfo{person}{Konstantinos
  Karanasos}.} \bibinfo{year}{2022}\natexlab{}.
\newblock \showarticletitle{End-to-End Optimization of Machine Learning
  Prediction Queries}. In \bibinfo{booktitle}{\emph{Proceedings of the 2022
  International Conference on Management of Data}} (Philadelphia, PA, USA)
  \emph{(\bibinfo{series}{SIGMOD '22})}. \bibinfo{publisher}{Association for
  Computing Machinery}, \bibinfo{address}{New York, NY, USA},
  \bibinfo{pages}{587–601}.
\newblock
\showISBNx{9781450392495}
\href{https://doi.org/10.1145/3514221.3526141}{doi:\nolinkurl{10.1145/3514221.3526141}}


\bibitem[Reagan(2018)]%
        {cosmos}
\bibfield{author}{\bibinfo{person}{Rob Reagan}.}
  \bibinfo{year}{2018}\natexlab{}.
\newblock \bibinfo{booktitle}{\emph{{Cosmos DB}}}.
\newblock \bibinfo{publisher}{Apress}, \bibinfo{address}{Berkeley, CA},
  \bibinfo{pages}{187--255}.
\newblock
\showISBNx{978-1-4842-2976-7}
\href{https://doi.org/10.1007/978-1-4842-2976-7_6}{doi:\nolinkurl{10.1007/978-1-4842-2976-7_6}}


\bibitem[Redis(2023)]%
        {redisHashtag}
\bibfield{author}{\bibinfo{person}{Redis}.} \bibinfo{year}{2023}\natexlab{}.
\newblock \bibinfo{booktitle}{\emph{Scaling with Redis Cluster}}.
\newblock
\urldef\tempurl%
\url{https://redis.io/docs/management/scaling/}
\showURL{%
\tempurl}


\bibitem[Robicquet et~al\mbox{.}(2016)]%
        {sdd}
\bibfield{author}{\bibinfo{person}{Alexandre Robicquet}, \bibinfo{person}{Amir
  Sadeghian}, \bibinfo{person}{Alexandre Alahi}, {and} \bibinfo{person}{Silvio
  Savarese}.} \bibinfo{year}{2016}\natexlab{}.
\newblock \showarticletitle{{Learning Social Etiquette: Human Trajectory
  Understanding In Crowded Scenes}}. In \bibinfo{booktitle}{\emph{Computer
  Vision -- ECCV 2016}}, \bibfield{editor}{\bibinfo{person}{Bastian Leibe},
  \bibinfo{person}{Jiri Matas}, \bibinfo{person}{Nicu Sebe}, {and}
  \bibinfo{person}{Max Welling}} (Eds.). \bibinfo{publisher}{Springer
  International Publishing}, \bibinfo{address}{Cham},
  \bibinfo{pages}{549--565}.
\newblock
\showISBNx{978-3-319-46484-8}


\bibitem[Shaowang et~al\mbox{.}(2021)]%
        {edgeserve}
\bibfield{author}{\bibinfo{person}{Ted Shaowang}, \bibinfo{person}{Nilesh
  Jain}, \bibinfo{person}{Dennis~D. Matthews}, {and} \bibinfo{person}{Sanjay
  Krishnan}.} \bibinfo{year}{2021}\natexlab{}.
\newblock \showarticletitle{Declarative Data Serving: The Future of Machine
  Learning Inference on the Edge}.
\newblock \bibinfo{journal}{\emph{Proc. VLDB Endow.}} \bibinfo{volume}{14},
  \bibinfo{number}{11} (\bibinfo{date}{jul} \bibinfo{year}{2021}),
  \bibinfo{pages}{2555–2562}.
\newblock
\showISSN{2150-8097}
\href{https://doi.org/10.14778/3476249.3476302}{doi:\nolinkurl{10.14778/3476249.3476302}}


\bibitem[Sheng et~al\mbox{.}(2025)]%
        {rlhf}
\bibfield{author}{\bibinfo{person}{Guangming Sheng}, \bibinfo{person}{Chi
  Zhang}, \bibinfo{person}{Zilingfeng Ye}, \bibinfo{person}{Xibin Wu},
  \bibinfo{person}{Wang Zhang}, \bibinfo{person}{Ru Zhang},
  \bibinfo{person}{Yanghua Peng}, \bibinfo{person}{Haibin Lin}, {and}
  \bibinfo{person}{Chuan Wu}.} \bibinfo{year}{2025}\natexlab{}.
\newblock \showarticletitle{HybridFlow: A Flexible and Efficient RLHF
  Framework}. In \bibinfo{booktitle}{\emph{Proceedings of the Twentieth
  European Conference on Computer Systems}} \emph{(\bibinfo{series}{EuroSys
  '25})}. \bibinfo{pages}{1279–1297}.
\newblock
\href{https://doi.org/10.1145/3689031.3696075}{doi:\nolinkurl{10.1145/3689031.3696075}}


\bibitem[Song et~al\mbox{.}(2023)]%
        {cascade_arxiv}
\bibfield{author}{\bibinfo{person}{Weijia Song}, \bibinfo{person}{Thiago
  Garrett}, \bibinfo{person}{Yuting Yang}, \bibinfo{person}{Mingzhao Liu},
  \bibinfo{person}{Edward Tremel}, \bibinfo{person}{Lorenzo Rosa},
  \bibinfo{person}{Andrea Merlina}, \bibinfo{person}{Roman Vitenberg}, {and}
  \bibinfo{person}{Ken Birman}.} \bibinfo{year}{2023}\natexlab{}.
\newblock \bibinfo{title}{Cascade: A Platform for Delay-Sensitive Edge
  Intelligence}.
\newblock
\showeprint[arxiv]{2311.17329}~[cs.OS]


\bibitem[Sreekanti et~al\mbox{.}(2020)]%
        {cloudburst}
\bibfield{author}{\bibinfo{person}{Vikram Sreekanti},
  \bibinfo{person}{Chenggang Wu}, \bibinfo{person}{Xiayue~Charles Lin},
  \bibinfo{person}{Johann Schleier-Smith}, \bibinfo{person}{Joseph~E.
  Gonzalez}, \bibinfo{person}{Joseph~M. Hellerstein}, {and}
  \bibinfo{person}{Alexey Tumanov}.} \bibinfo{year}{2020}\natexlab{}.
\newblock \showarticletitle{Cloudburst: Stateful Functions-as-a-Service}.
\newblock \bibinfo{journal}{\emph{Proc. VLDB Endow.}} \bibinfo{volume}{13},
  \bibinfo{number}{12} (\bibinfo{date}{jul} \bibinfo{year}{2020}),
  \bibinfo{pages}{2438–2452}.
\newblock
\showISSN{2150-8097}
\href{https://doi.org/10.14778/3407790.3407836}{doi:\nolinkurl{10.14778/3407790.3407836}}


\bibitem[Storm(2023)]%
        {stormGrouping}
\bibfield{author}{\bibinfo{person}{Apache Storm}.}
  \bibinfo{year}{2023}\natexlab{}.
\newblock \bibinfo{booktitle}{\emph{Concepts}}.
\newblock
\urldef\tempurl%
\url{https://storm.apache.org/releases/current/Concepts.html}
\showURL{%
\tempurl}


\bibitem[Wang et~al\mbox{.}(2019)]%
        {hyperscan}
\bibfield{author}{\bibinfo{person}{Xiang Wang}, \bibinfo{person}{Yang Hong},
  \bibinfo{person}{Harry Chang}, \bibinfo{person}{KyoungSoo Park},
  \bibinfo{person}{Geoff Langdale}, \bibinfo{person}{Jiayu Hu}, {and}
  \bibinfo{person}{Heqing Zhu}.} \bibinfo{year}{2019}\natexlab{}.
\newblock \showarticletitle{Hyperscan: A Fast Multi-pattern Regex Matcher for
  Modern {CPUs}}. In \bibinfo{booktitle}{\emph{16th USENIX Symposium on
  Networked Systems Design and Implementation (NSDI 19)}}.
  \bibinfo{publisher}{USENIX Association}, \bibinfo{address}{Boston, MA},
  \bibinfo{pages}{631--648}.
\newblock
\showISBNx{978-1-931971-49-2}
\urldef\tempurl%
\url{https://www.usenix.org/conference/nsdi19/presentation/wang-xiang}
\showURL{%
\tempurl}


\bibitem[Yu et~al\mbox{.}(2021)]%
        {chukonu}
\bibfield{author}{\bibinfo{person}{Bowen Yu}, \bibinfo{person}{Guanyu Feng},
  \bibinfo{person}{Huanqi Cao}, \bibinfo{person}{Xiaohan Li},
  \bibinfo{person}{Zhenbo Sun}, \bibinfo{person}{Haojie Wang},
  \bibinfo{person}{Xiaowei Zhu}, \bibinfo{person}{Weimin Zheng}, {and}
  \bibinfo{person}{Wenguang Chen}.} \bibinfo{year}{2021}\natexlab{}.
\newblock \showarticletitle{Chukonu: A Fully-Featured High-Performance Big Data
  Framework That Integrates a Native Compute Engine into Spark}.
\newblock \bibinfo{journal}{\emph{Proc. VLDB Endow.}} \bibinfo{volume}{15},
  \bibinfo{number}{4} (\bibinfo{date}{dec} \bibinfo{year}{2021}),
  \bibinfo{pages}{872–885}.
\newblock
\showISSN{2150-8097}
\href{https://doi.org/10.14778/3503585.3503596}{doi:\nolinkurl{10.14778/3503585.3503596}}


\bibitem[Yu et~al\mbox{.}(2023)]%
        {pheromone}
\bibfield{author}{\bibinfo{person}{Minchen Yu}, \bibinfo{person}{Tingjia Cao},
  \bibinfo{person}{Wei Wang}, {and} \bibinfo{person}{Ruichuan Chen}.}
  \bibinfo{year}{2023}\natexlab{}.
\newblock \showarticletitle{{Following the Data, Not the Function: Rethinking
  Function Orchestration in Serverless Computing}}. In
  \bibinfo{booktitle}{\emph{20th USENIX Symposium on Networked Systems Design
  and Implementation (NSDI 23)}}. \bibinfo{publisher}{USENIX Association},
  \bibinfo{address}{Boston, MA}, \bibinfo{pages}{1489--1504}.
\newblock
\showISBNx{978-1-939133-33-5}
\urldef\tempurl%
\url{https://www.usenix.org/conference/nsdi23/presentation/yu}
\showURL{%
\tempurl}


\bibitem[Zaharia et~al\mbox{.}(2012)]%
        {rdd}
\bibfield{author}{\bibinfo{person}{Matei Zaharia}, \bibinfo{person}{Mosharaf
  Chowdhury}, \bibinfo{person}{Tathagata Das}, \bibinfo{person}{Ankur Dave},
  \bibinfo{person}{Justin Ma}, \bibinfo{person}{Murphy McCauly},
  \bibinfo{person}{Michael~J. Franklin}, \bibinfo{person}{Scott Shenker}, {and}
  \bibinfo{person}{Ion Stoica}.} \bibinfo{year}{2012}\natexlab{}.
\newblock \showarticletitle{Resilient Distributed Datasets: A {Fault-Tolerant}
  Abstraction for {In-Memory} Cluster Computing}. In
  \bibinfo{booktitle}{\emph{9th USENIX Symposium on Networked Systems Design
  and Implementation (NSDI 12)}}. \bibinfo{publisher}{USENIX Association},
  \bibinfo{address}{San Jose, CA}, \bibinfo{pages}{15--28}.
\newblock
\showISBNx{978-931971-92-8}
\urldef\tempurl%
\url{https://www.usenix.org/conference/nsdi12/technical-sessions/presentation/zaharia}
\showURL{%
\tempurl}


\bibitem[Zhang et~al\mbox{.}(2017)]%
        {videostorm}
\bibfield{author}{\bibinfo{person}{Haoyu Zhang}, \bibinfo{person}{Ganesh
  Ananthanarayanan}, \bibinfo{person}{Peter Bodik}, \bibinfo{person}{Matthai
  Philipose}, \bibinfo{person}{Paramvir Bahl}, {and}
  \bibinfo{person}{Michael~J. Freedman}.} \bibinfo{year}{2017}\natexlab{}.
\newblock \showarticletitle{Live Video Analytics at Scale with Approximation
  and {Delay-Tolerance}}. In \bibinfo{booktitle}{\emph{14th USENIX Symposium on
  Networked Systems Design and Implementation (NSDI 17)}}.
  \bibinfo{publisher}{USENIX Association}, \bibinfo{address}{Boston, MA},
  \bibinfo{pages}{377--392}.
\newblock
\showISBNx{978-1-931971-37-9}
\urldef\tempurl%
\url{https://www.usenix.org/conference/nsdi17/technical-sessions/presentation/zhang}
\showURL{%
\tempurl}


\bibitem[Zhou et~al\mbox{.}(2022b)]%
        {osnet}
\bibfield{author}{\bibinfo{person}{Kaiyang Zhou}, \bibinfo{person}{Yongxin
  Yang}, \bibinfo{person}{Andrea Cavallaro}, {and} \bibinfo{person}{Tao
  Xiang}.} \bibinfo{year}{2022}\natexlab{b}.
\newblock \showarticletitle{Learning Generalisable Omni-Scale Representations
  for Person Re-Identification}.
\newblock \bibinfo{journal}{\emph{IEEE Transactions on Pattern Analysis and
  Machine Intelligence}} \bibinfo{volume}{44}, \bibinfo{number}{9}
  (\bibinfo{year}{2022}), \bibinfo{pages}{5056--5069}.
\newblock
\href{https://doi.org/10.1109/TPAMI.2021.3069237}{doi:\nolinkurl{10.1109/TPAMI.2021.3069237}}


\bibitem[Zhou et~al\mbox{.}(2022a)]%
        {db4ai2}
\bibfield{author}{\bibinfo{person}{Xuanhe Zhou}, \bibinfo{person}{Chengliang
  Chai}, \bibinfo{person}{Guoliang Li}, {and} \bibinfo{person}{Ji Sun}.}
  \bibinfo{year}{2022}\natexlab{a}.
\newblock \showarticletitle{Database Meets Artificial Intelligence: A Survey}.
\newblock \bibinfo{journal}{\emph{IEEE Transactions on Knowledge and Data
  Engineering}} \bibinfo{volume}{34}, \bibinfo{number}{3}
  (\bibinfo{year}{2022}), \bibinfo{pages}{1096--1116}.
\newblock
\href{https://doi.org/10.1109/TKDE.2020.2994641}{doi:\nolinkurl{10.1109/TKDE.2020.2994641}}


\end{thebibliography}


\end{document}